\documentclass[fleqn,usenatbib,useAMS]{mnras}

\DeclareMathAlphabet{\mathsc}{OT1}{cmr}{m}{sc}
\def\testbx{bx}%
\DeclareRobustCommand{\ion}[2]{%
\relax\ifmmode
\ifx\testbx\f@series
{\mathbf{#1\,\mathsc{#2}}}\else
{\mathrm{#1\,\mathsc{#2}}}\fi
\else\textup{#1\,{\mdseries\textsc{#2}}}%
\fi}

\usepackage{graphicx}
\graphicspath{{figures/}}
\usepackage{xcolor}
\usepackage[percent]{overpic}
\usepackage{mathptmx}
\usepackage{anyfontsize}
\usepackage{t1enc}
\usepackage{url}
\usepackage{lineno}
\title{The Center-to-Limb Variation of Non-Thermal Velocities using IRIS Si IV}
\author[Yamini K. Rao, Giulio Del Zanna, and Helen E. Mason]{Yamini K. Rao$^{1}$\thanks{E-mail: yy496@cam.ac.uk}, Giulio Del Zanna$^{1}$\thanks{E-mail: gd232@cam.ac.uk}, and Helen E. Mason $^{1}$\\
$^{1}$DAMTP, University of Cambridge, Wilberforce Rd, Cambridge CB3 0WA, United Kingdom
}
\begin{document}

\date{27 July 2021}

\pagerange{\pageref{firstpage}--\pageref{lastpage}} \pubyear{2021}

\maketitle

\label{firstpage}

\begin{abstract}

We study the non-thermal velocities in the quiet-sun using various high spatial, temporal, and spectral resolution observations from the Interface Region Imaging Spectrograph (IRIS). We focus our analysis on the transition region using the optically thin line (\ion{Si}{iv} 1393.7 \AA), and select line profiles that are nearly Gaussian. We find evidence of a centre-to-limb variation using different observations having different exposure times, ranging from 5-30 s. 
The distribution of non-thermal velocities close to the limb are observed to peak around 20 km s$^{-1}$ while the disc observations show a peak around 15 km s$^{-1}$. The distributions are also different. The overall variation in the non-thermal velocities are correlated with the intensity of the line, as found previously. The on-disc velocities are smaller than most previous observations. In general, we find that the  non-thermal velocities are independent of the selected exposure times. The \ion{Si}{iv} lines didn't seem to exhibit any significant opacity effects. We conclude that these Doppler motions are mostly transverse to the radial direction. The possibility of swaying/torsional motions leading to such variations are validated from these IRIS observations.

\end{abstract}

\section{Introduction}

At coronal temperatures, the Sun can be distinctly categorised into several different 
regions, such as Active Regions (ARs), Quiet Sun (QS) and Coronal Holes (CH). 
At transition region (TR) temperatures, the structures appear almost the same in the QS and CH. Such structures are organised in a super-granular network pattern. In the cell centres the intensities are lower and less variable, while at the boundaries there is more dynamic activity. 
The nature of the TR and its role in connecting the chromosphere with the corona is still a topic of debate in the literature. It is likely that at least part of the TR emission which we observe must be related to magnetic structures threading the chromosphere and reaching the corona. Observationally, the QS TR lines show a net redshift, which varies with their formation temperature. This represents bulk motions of radiatively cooling plasma. 

The QS TR  also exhibits unresolved motions which are characterised by excess broadening in the profiles of the spectral lines, in addition to the thermal broadening and the instrumental broadening. Such broadenings are the focus of this paper.
These plasma flows transfer mass and energy between the chromosphere and the corona
and are likely to be intimately related to the (still unknown) physical processes occurring in the TR. For these reasons, many previous studies have focused on observations of non-thermal broadenings. 

In terms of the full-width-at-half-maximum (FWHM) of a line, the contribution from thermal motions, instrumental broadening, and non-thermal motions,
assuming Gaussian profiles, is:
\begin{equation}
\centering
{\mathrm{FWHM}^{2} = {w_{I}}^{2} + {w_{O}}^{2} = 
{w_{I}}^{2} + 4\,ln\,2  (\frac{\lambda_{o}}{c})^{2} \left[\frac{2kT_{i}}{M} + \xi^{2} \right]  }
\end{equation}
where $w_{I}$ is the instrumental FWHM and $w_{O}$ is the observed FWHM once the instrumental width is subtracted. The first component of $w_{O}$ is due to the thermal broadening of the ions (assuming a Maxwell-Boltzmann distribution of velocities) where $T_{i}$, $M$, $\lambda_{o}$, $c$ are the temperature of the ion, the mass of the ion, the rest wavelength and speed of light respectively. Here, $\xi$ is the most probable non-thermal velocity (NTV) for a Maxwellian. The non-thermal velocity discussed in this paper is in effect the non-thermal width observed as an excess broadening in the spectral profile due to unresolved motions and is obtained using the measured FWHM of the Gaussian. In earlier literature, the non-thermal velocity is sometimes defined using different widths (e.g. FWHM, Gaussian sigma, w$_{1/e}$). We measured the FWHM to obtain the NTV.  The NTV values in  Table \ref{table_prev} were also obtained from measurements of the FWHM of the lines.

 
\begin{table*}
\centering
\begin{tabular}{c c c c c c}
\hline
No. & Instrument          & Spectral Line     &  Location & Exposure time (s)   &  NTV (km s$^{-1}$)   \\
 \hline
1. & Echelle Spectrograph 	& \ion{C}{iv}               & Disc         & 47            & 19 \\
2. & Echelle Spectrograph 	& \ion{Si}{iv}              & Disc         & 29            & 14 \\
3. & Skylab S082-B NRL 		& \ion{Si}{iv}; \ion{C}{iv} & Limb         & -             & 22-25  \\
4. & Skylab S082-B NRL 		& \ion{Si}{iv}; \ion{C}{iv} & Off Limb     & $>$ 600       & 32  \\
5. & Skylab S082-B NRL 	    & \ion{Si}{iv}; \ion{C}{iv} & Off Limb     & -             & 37-43  \\
6. & HRTS		            & \ion{C}{iv}               & Limb         &  3            & 10-25  \\
7. & HRTS		            & \ion{C}{iv}               & Disc         &  1-20         & 22  \\
8. & HRTS                   & \ion{Si}{iv}              & Disc         & 1-20          & 28 \\
9. & SUMER                  & \ion{C}{iv}               & Disc         & 100-300       & 25 \\
10. & SUMER                 & \ion{Si}{iv}              & Disc         & 100-300       & 23\\
11.& SUMER                  & \ion{Si}{iv}              & Disc         & 180           & 20-30 \\
12.& SUMER                  & \ion{C}{iv}               & Disc         & 15            & 25 \\
\end{tabular}
\caption{List of various previous measurements of NTVs obtained from measurements of the FWHM in the transition region using \ion{Si}{iv} and \ion{C}{iv} lines from different instruments as mentioned 1,2. \citet{1975MNRAS.171..697B} 3,4,5. \citet{1978ApJ...226..698M} 6. \citet{1989SoPh..123...41D} 7,8. \citet{1993SoPh..144..217D} 9,10. \citet{1998ApJ...505..957C} 11. \citet{2005ApJ...623..540A}; 12. \citet{1999ApJ...516..490P} with modifications following \citet{2008ApJ...673L.219M}.
}
\label{table_prev}
\end{table*}
The Interface Region Imaging Spectrograph (IRIS; \citealt{2014SoPh..289.2733D}) provides simultaneous imaging and spectral data in the FUV (1331.7 \AA\ $-$ 1358.4 \AA\ and 1389.0 \AA $-$ 1407.0 \AA) and NUV (2782.7 \AA\ $-$ 2835.1 \AA). The spectral lines observed in this range cover the photosphere, chromosphere, TR and coronal temperatures. We primarily focus on the Si IV doublet (1389.0 \AA\ $-$ 1407.0 \AA)  as these are the strongest TR lines.

IRIS is an excellent instrument to measure NTVs compared to most previous instruments for four main reasons: 
1)  it has a very narrow instrumental broadening, equivalent to about 6~km/s (0.03~\AA) for the \ion{Si}{ iv} lines;
2) there is a good sampling across the line profiles (about 20 IRIS pixels sample the line profiles);
3) the large collecting area means that short exposures (10~s or less) can be achieved. This
is much better than the long exposures needed for most previous instruments;
4) IRIS  has a much higher spatial resolution (slit width 0.33\arcsec) than previous instruments in the UV (at best around 1\arcsec). 
As the NTVs are likely due to a superposition of many flows within the resolution element, and as we observe that the TR lines are very dynamic,
one would expect to find different NTVs by increasing the spatial resolution and/or lowering the exposure times. 

The spatial resolution effect was studied using IRIS  \ion{Si}{iv} by \citet{2015ApJ...799L..12D}. They considered three different regions (AR, QS, and CH) and found that the peak of the distribution of the NTVs is invariant with the spatial resolution, but exhibits variations in the wings. This result is also analogous to what \cite{2016ApJ...827...99T} found for the hotter coronal emission in \ion{Fe}{xii}. 
In this paper, we extend their study by analysing several IRIS observational datasets for the QS at different locations using the \ion{Si}{iv} lines,
and we also measure the NTVs for different exposure times.

There are puzzling differences in the NTV measurements across the literature obtained with previous instrumentation. To understand such differences it is necessary to compare the different instrumental and observational parameters. An overview is provided in Section~2, focused on the \ion{C}{iv} and \ion{Si}{iv} measurements of the optically allowed doublets in the UV (\ion{C}{iv} is also considered as it has a formation temperature close to that of \ion{Si}{iv}).

In section~3 we present the details of the
observational data, the results and their interpretation. 
In the last section, the results are summarised and conclusions are presented. 
\section{Previous observations}

For a short overview of previous observations of non-thermal widths see the {\it Solar Physics Living Review} by  \cite{2018LRSP...15....5D}. Here, we provide key details of some of the previous observations of NTVs in \ion{Si}{iv} and \ion{C}{iv} lines, in chronological order. A short summary is given in Table~\ref{table_prev}. 
We  note that some authors such as (\citealt{1973A&A....22..161B}, \citealt{1975MNRAS.171..697B}) tabulated the values of the root-mean-square velocity v$_{\mathrm rms}= \sqrt{<v^2>}$, which is related to the the most probable non-thermal speed $\xi$ ( v$_{\mathrm rms}$ =  $\sqrt[]{3/2}$ $\xi$). We have converted the Boland et al. values to our definition of NTV and added them to Table \ref{table_prev}.

\begin{center}
\begin{figure*}
\includegraphics[scale=0.4,angle=90,width=15cm,height=12cm,keepaspectratio]{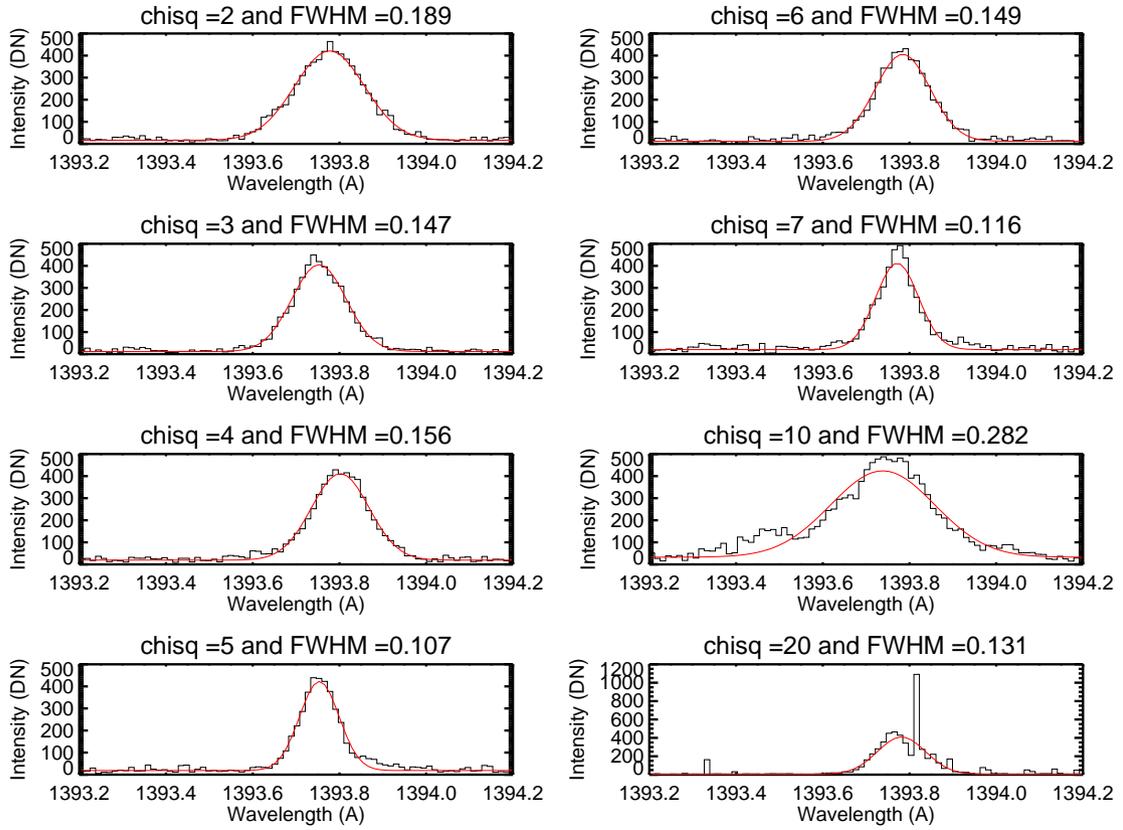}
\caption{Spectral fitting for a sample of 1\arcsec\ IRIS pixels, having peak data numbers around 400 and different chi-squared values. The red line is the Gaussian fitting plus a linear background. The chi-square and the FWHM in \AA\ are shown in each plot.}
\label{spec_fit}
\end{figure*}
\end{center}
Excellent measurements of the line widths in UV TR lines were obtained with an Echelle spectrograph, with a maximum resolution of 0.026~\AA\ FWHM, flown twice on a Skylark sounding rocket (\citealt{1973A&A....22..161B}, \citealt{1975MNRAS.171..697B} and references therein). The instrumental width was measured in-flight from the Fraunhofer lines. During the first flight, a 100~s exposure on the quiet Sun indicated an average FWHM of 0.24~\AA\ for the \ion{C}{iv} 1548~\AA\ line, resulting in an intrinsic width of 0.17$\pm0.08$. The large uncertainty was due to the larger instrumental FWHM (0.12~\AA) for that exposure. During the second flight, only a 40\arcsec\ portion of the slit, pointed about 10\arcmin\ from Sun centre,
could be used. The exposures were 47 and 29~s. The \ion{C}{iv} 1551~\AA\ line had an average  FWHM of 0.207~\AA, resulting in an intrinsic width of 0.19$\pm0.01$. Assuming a temperature of
1$\times$10$^5$ K, this is equivalent to a NTV (FWHM) of 19 km s$^{-1}$. The stronger of the
\ion{Si}{iv} doublet, at 1394~\AA, had a width of 0.138~\AA\ and an intrinsic width of 0.12$\pm0.01$ (with an estimated instrumental
FWHM of 0.068~\AA). Assuming a temperature of 6.3 $\times$10$^4$ K, this is equivalent to
a NTV (FWHM) of 14 km s$^{-1}$, i.e. close to that of \ion{C}{iv}, as one would expect. 

\begin{center}
\begin{figure*}
\includegraphics[scale=0.4,angle=90,width=15cm,height=12cm,keepaspectratio]{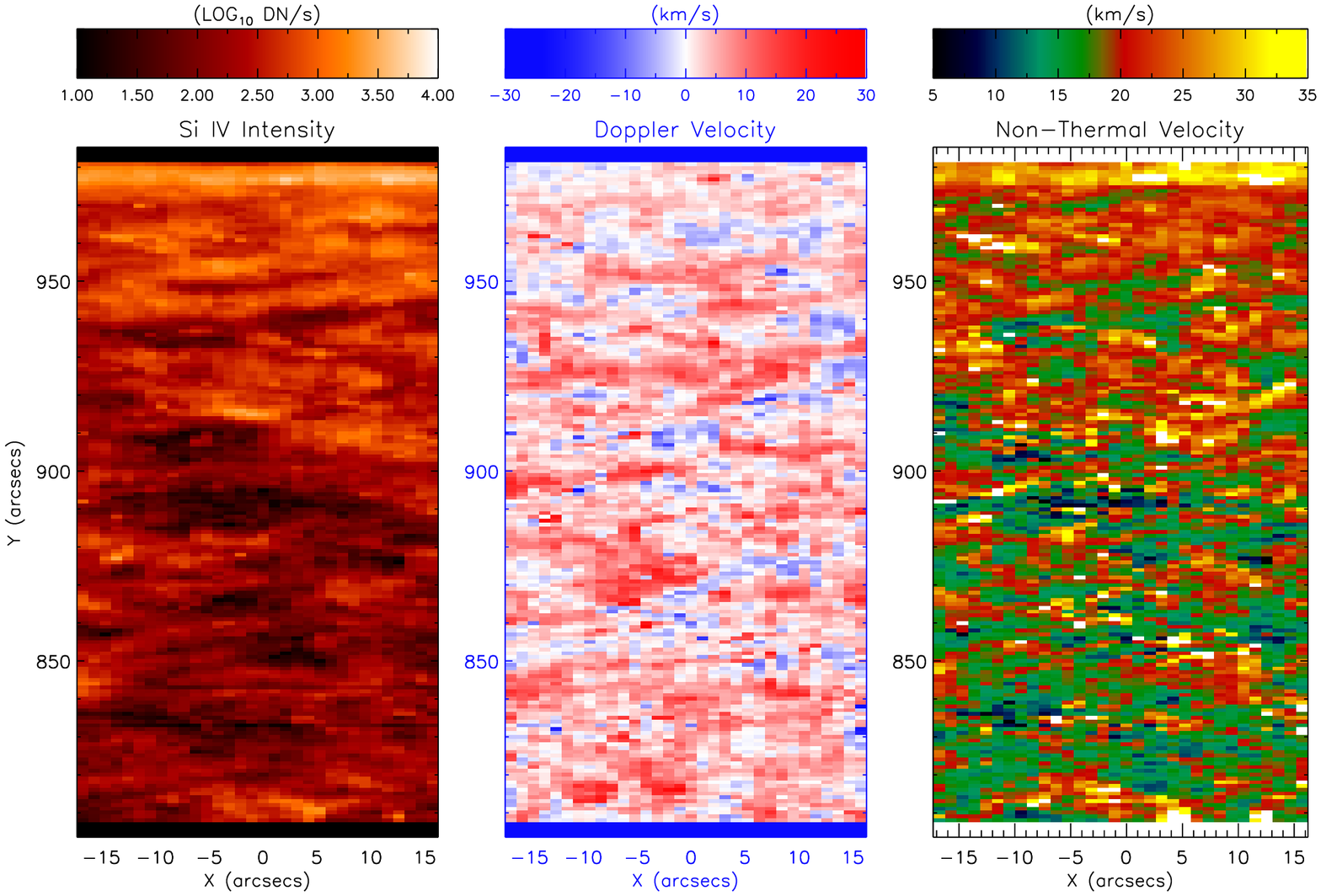}
\caption{Parametric plots of the QS observed on the 22nd January, 2014 close to the northern limb having an exposure time of 15 s. The left panel shows the intensity of the IRIS \ion{Si}{iv} 1393.75 \AA\ line. The middle and right panels indicate the corresponding Doppler velocities and Non-thermal velocities. The colour bars with the scales are shown above the plots. X and Y are distances from Sun centre in arc seconds.}
\label{fig1}
\end{figure*}
\end{center}

\begin{figure}
\includegraphics[scale=0.5,angle=90,width=9cm,keepaspectratio]{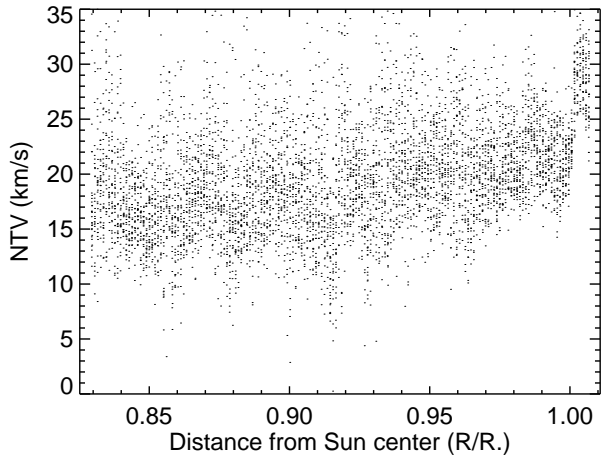}
\caption{The distribution of non-thermal velocity along radial distance from the Sun's center for the observation shown in Fig. \ref{fig1}}
\label{fig1extra}
\end{figure}

\begin{center}
\begin{figure*}
\mbox{
\includegraphics[scale=0.4,angle=90,width=8cm,height=10cm,keepaspectratio]{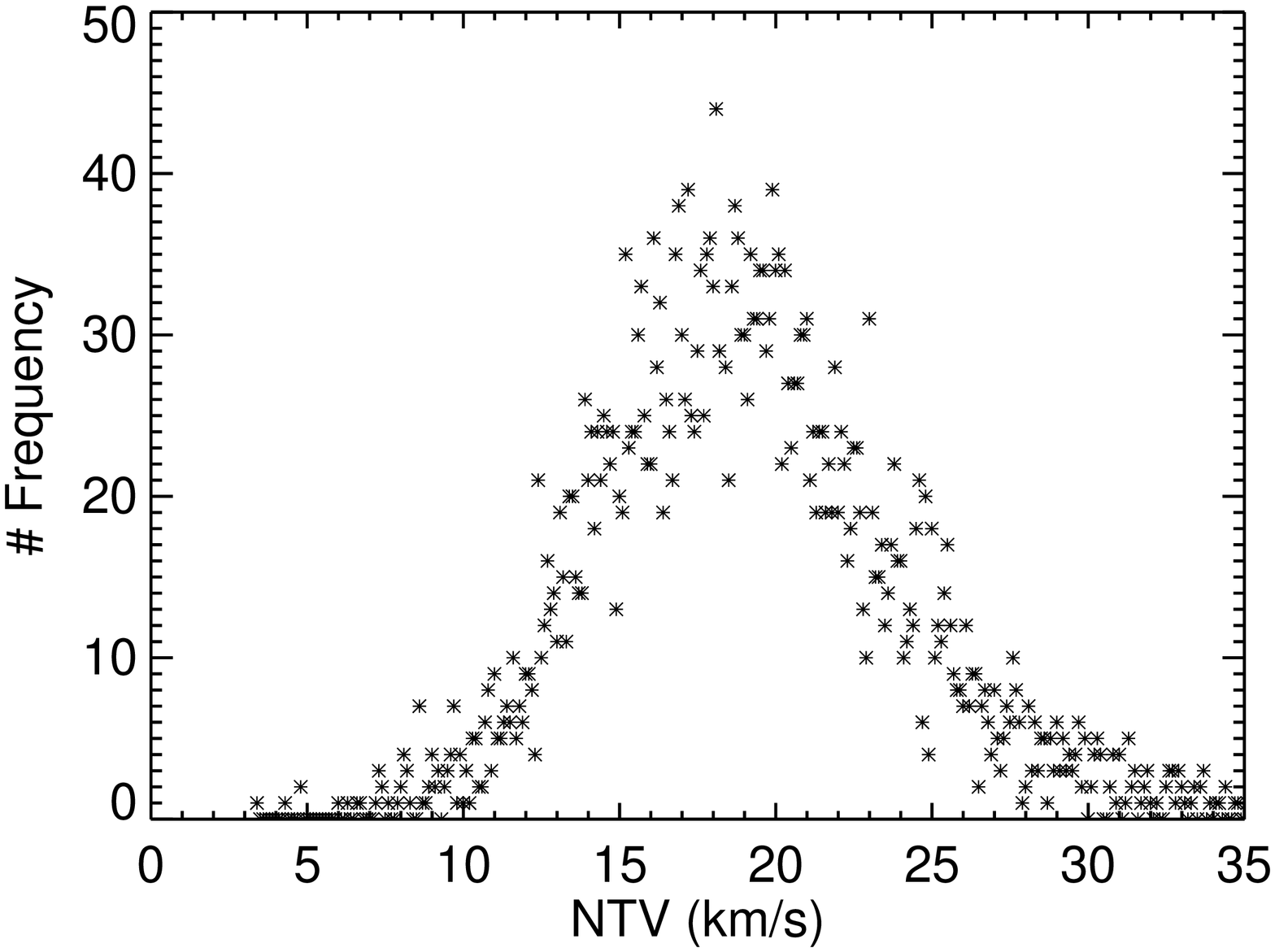}

\includegraphics[scale=0.4,angle=90,width=8cm,height=10cm,keepaspectratio]{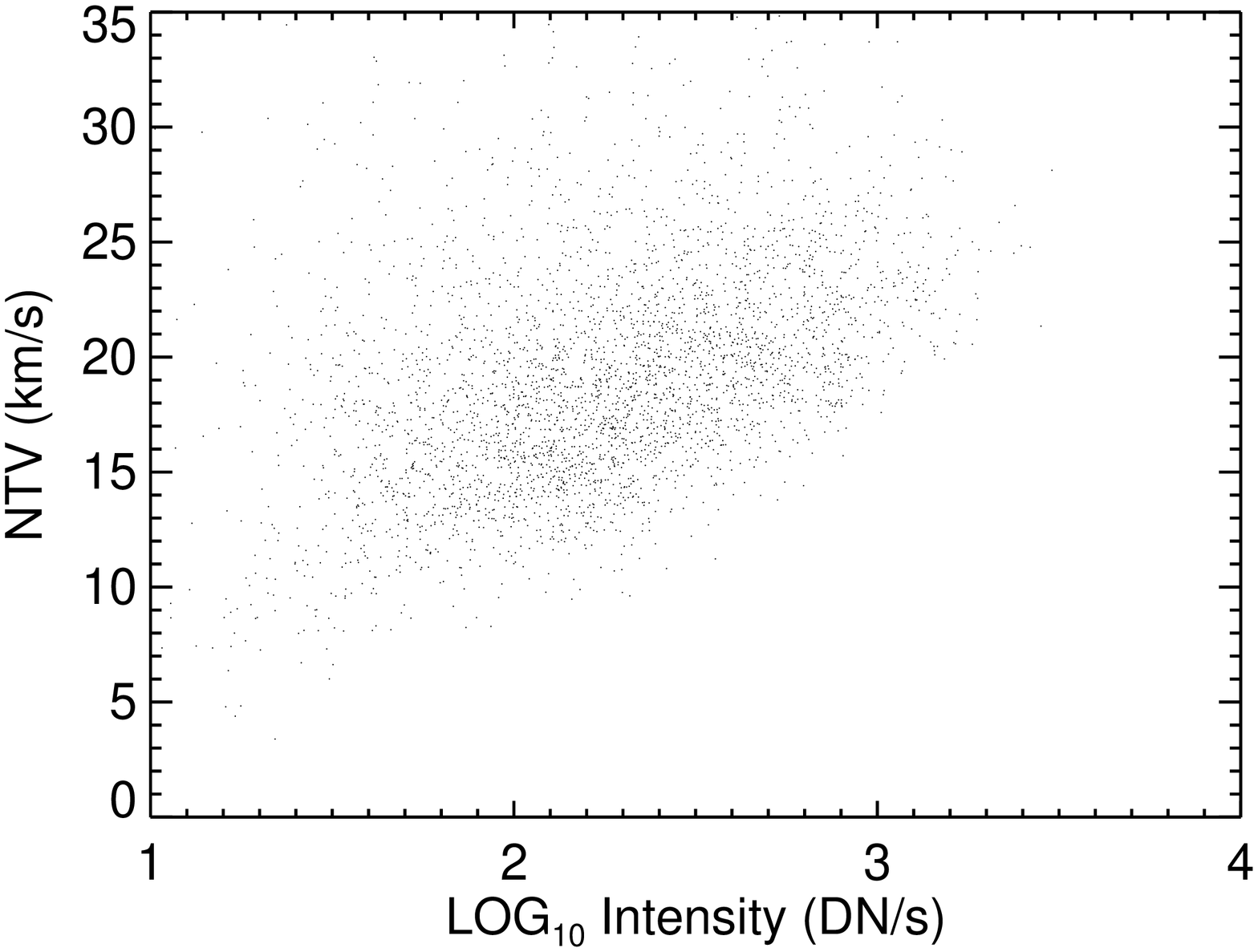}
}
\caption{Left panel: The distribution of non-thermal velocity for the QS observed on the 22nd January 2014 for which the parametric plot is shown in Fig. \ref{fig1}. Right panel: A scatter plot of NTVs as a function of the intensity of the \ion{Si}{iv} 1393.75 \AA\ spectral line.}
\label{fig2}
\end{figure*}
\end{center}

The Skylab observations from the Naval Research Laboratory (NRL) S082B instrument with an 2x60\arcsec\ slit provided many  NTV measurements for TR lines. One limitation was that the instrument was astigmatic, with no spatial resolution along the slit. Also observations were usually carried out at the solar limb, rather than on the disc. The instrument resolution (FWHM) was estimated to be about 0.06~\AA. Generally, the Skylab NTVs of \ion{Si}{iv} and \ion{C}{iv} are larger (around 22-25 km s$^{-1}$) than the Boland measurements,
although the differences were not noted (cf. \citealt{1977ApJS...33..101D},
\citealt{1978ApJ...226..698M}).
Indeed, in the literature, it has been generally assumed that the NTVs are isotropic, as it was thought that centre-to-limb variations were not present \citep[see,e.g.][]{1978ApJ...226..698M}.
Also, the Skylab observations showed that lines such as \ion{Si}{iv} and \ion{C}{iv} have NTVs  that increase with height above the limb, reaching 32~km s$^{-1}$ at 12\arcsec\ above the limb and 37 - 43~km s$^{-1}$ at 20\arcsec\
(cf. \citealt{1979A&A....73..361M}). However, exposures longer than 600~s were required for the observations  12\arcsec\ above the limb, so the 
increase could well be due to a superposition of motions during these long exposures. It was thought that some excess broadenings in the lines could be due to opacity effects, although line ratios indicated that opacity was only present near the limb \citep[see, e.g.][]{1980MNRAS.193..947D}.

Differences in the NTVs of allowed and inter-combination lines were noted in the Skylab spectra, and there is an extended, early literature on this issue.  However, IRIS observations of the \ion{Si}{iv} allowed lines and the \ion{O}{iv} inter-system lines have indicated that this is not the case
(see \citealt{2016A&A...594A..64P}, \citealt{2016ApJ...832...77D},
\citealt{2017SoPh..292..100D}, and reference therein), at least for these transitions. We note however that the wings of the forbidden lines 
are always difficult to measure as these lines are intrinsically much weaker than the allowed transitions \citep[see,e.g.][]{1993SoPh..144..217D}. 

The Skylab observations were very important in showing that the excess broadening is a real effect, present in the quiet sun and coronal holes, as NTVs of just a few km s$^{-1}$
were observed above sunspots (e.g. \citealt{1976ApJ...210..836C}) and in prominences (e.g. \citealt{1977ApJ...216L.119F}).

The High Resolution Telescope and Spectrometer (HRTS) produced many excellent results from the sounding rocket program. The HRTS stigmatic slit was very long, from Sun centre to slightly off-limb. The HRTS instrument was also flown on the Spacelab-2 Shuttle mission, returning a lot of data. The instrumental FWHM was estimated to be about 0.06~\AA, i.e. similar to that of the 
Skylab NRL S082B instrument.
\citet{1989SoPh..123...41D} used a series of exposures, from 3 to 350 s, to obtain quiet Sun spectra over a range of locations, finding the distribution of NTVs for \ion{C}{iv} (assuming  1$\times$10$^5$ K) between 10 and 25 km s$^{-1}$ (cf their Fig.11), with an average of 16 km s$^{-1}$, i.e. close to the Boland et al. results.

In a later paper, \citet{1993SoPh..144..217D} used HRTS observations from the first sounding rocket flight, with exposures ranging from
1 to 20~s. Those exposures were summed, selecting quiet Sun regions, to obtain an average NTV for the \ion{C}{iv} 1548~\AA\ line of 28 km s$^{-1}$.
For the \ion{Si}{iv} 1394~\AA, an average NTV of 22 km s$^{-1}$ was found. Such values are close to those in the previous literature, as listed in their paper, except to those of Boland et al.,
and those of the HRTS Spacelab-2 flight, which are significantly lower. 

It is well known that many line profiles are non-Gaussian, with extended broad wings especially in the explosive events. With a double Gaussian fit, \citet{1993SoPh..144..217D} found that the average of the NTV of the main component is about 15 km s$^{-1}$ for both lines, whilst the averaged width of the broad component is more than twice as large. The broad component is however usually very weak, a few percent the main component. By definition, the NTV is obtained assuming Gaussian profiles, hence the selection of near-Gaussian profiles becomes an important issue. It is often not clear from previous literature if locations with non-Gaussian profiles were excluded from the data analysed. Also, the precise location of the samples are often not given.

All the above observations were somewhat limited by the relatively few spatial locations considered. The SoHO SUMER instrument improved on this by providing a lot of measurements. 
\citet{1998ApJ...505..957C} estimated an instrumental FWHM of 2.3 detector pixels,
equivalent to 0.095~\AA\ around 1500~\AA, i.e.
about 12 km s$^{-1}$ (note that the dispersion changes slightly with wavelength).
The NTV of the narrow photospheric \ion{O}{i} 1355.6~\AA\ line resulted in 7 km s$^{-1}$, significantly larger than the value measured by the Skylab NRL S082B and HRTS instruments, which was about 4 km s$^{-1}$. It is therefore likely that the SUMER instrumental FWHM has been
under-estimated. Assuming that the Skylab and HRTS measurements for this line are correct, the FWHM around 1500~\AA\ would then result in 0.11~\AA, i.e. nearly a factor of two worse than the above-mentioned Echelle spectrograph and HRTS.

Having a larger instrumental width makes measurements of the NTVs more difficult. This, together with the fact that most SUMER observations had very long exposures (100 $-$ 300~s), could be the reason why most SUMER analyses produced NTV values in the \ion{C}{iv} and \ion{Si}{iv} significantly higher than 
previous analyses. For example, \citet{1998ApJ...505..957C} measured average NTVs for the \ion{Si}{iv} and \ion{C}{iv} lines
of 23 and 25 km s$^{-1}$, assuming ion temperatures of 7, 10 $\times$10$^4$ K
respectively.

\cite{1998A&A...337..287E} reported centre to limb variations in the upper chromosphere and transition region using various spectral lines from SUMER. However, close inspection of their results (see their Table 2) shows no significant differences in the NTVs between disc centre and near the limb, with e.g. a NTV of 9 km s$^{-1}$ for \ion{S}{iv}. Higher NTVs were only observed at the limb, as previously measured by Skylab. 

A special SUMER observing sequence scanning the whole Sun 
with very short exposure times (15 s) was analysed by 
\citet{1999ApJ...516..490P} to find an averaged NTV of 15 km s$^{-1}$ in \ion{C}{iv}, i.e. much lower than all other SUMER analyses. 
However, \cite{2008ApJ...673L.219M} later pointed that \citet{1999ApJ...516..490P}
applied an incorrect instrumental width, underestimating the NTV, which should have been 25 km s$^{-1}$, i.e. similar to the other published results from SUMER.
\citet{1999ApJ...516..490P} also reported a small decrease in the NTV 
towards the limb.
On the other hand, both 
\citet{2000A&A...356..335D} and \cite{2008ApJ...673L.219M} analysed the same observations but reported instead a small increase of a few km s$^{-1}$ towards the limb (we note that the values in those papers are not directly comparable as different velocities are displayed).  

\citet{2005ApJ...623..540A} selected a QS region observed by SUMER  with the 300\arcsec\ slit and and exposure time of 180~s. They measured an NTV for \ion{Si}{iv} between 20 and 30 km s$^{-1}$. A pointing at Sun centre gave an averaged value of 25 km s$^{-1}$, whilst one closer to the limb gave a slightly higher value, 27 km s$^{-1}$. Profiles with strongly non-Gaussian shapes or with low counts were selected out in this case. 
Similar measurements were provided by \citet{2000A&A...357..743L}. However, all of these observations had long exposure times. \cite{2001A&A...374.1108P} analysed SUMER observations with 115 s
and found much larger NTVs, about 30 s$^{-1}$.

In conclusion, considering the issues with instrumental widths, we consider the 
Boland et al. results as the most accurate among those obtained before IRIS. The Skylab NRL S082B and HRTS instruments had a lower spectral resolution and generally provided larger NTVs. The Skylab NRL S082B results were mostly near the limb and lacked spatial resolution. The HRTS results from the Spacelab-2 mission are very close to those of Boland et al. 

The IRIS results shown by 
\citet{2015ApJ...799L..12D} indicated NTVs ranging between 10 and 20. Single Gaussians were fitted. These values are significantly lower than most previous observations, but are very close to those of Boland et al. 
\begin{center}
\begin{figure*}
\includegraphics[scale=0.4,angle=90,width=15cm,height=12cm,keepaspectratio]{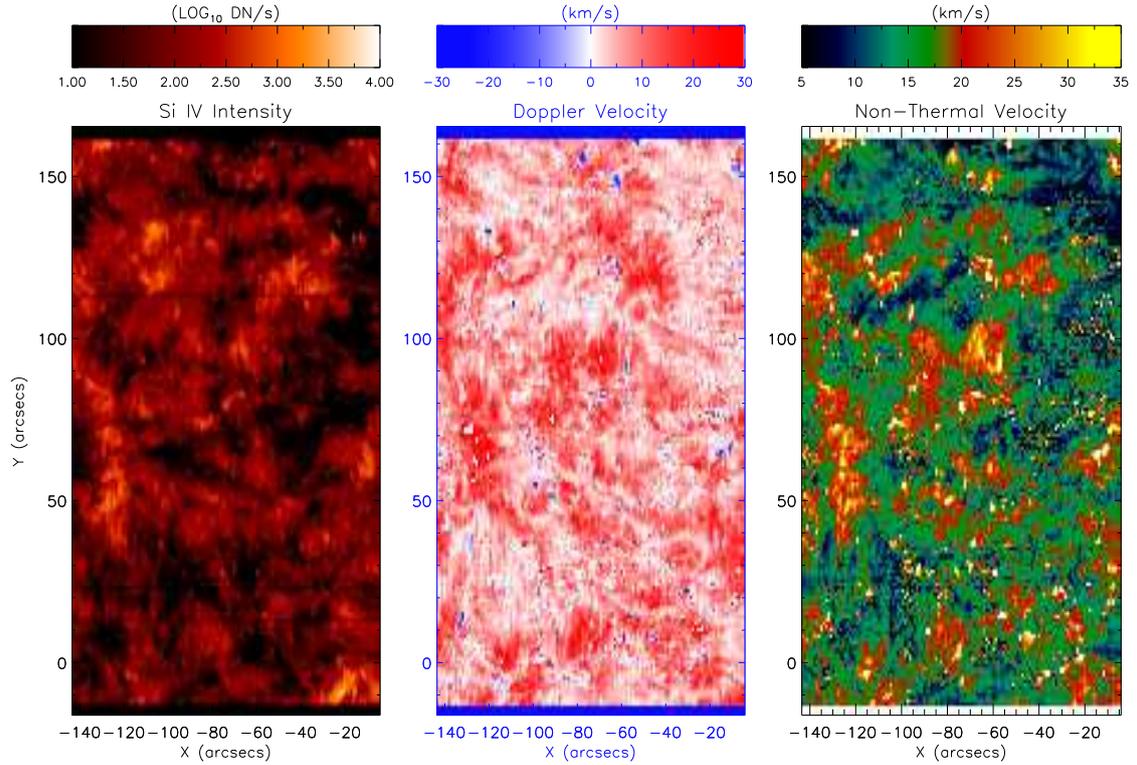}
\caption{Parametric plots of the QS observed on 25th Febraury 2014 near the disc center having an exposure time of 30~s (same as Fig.~\ref{fig1}). }
\label{3}
\end{figure*}
\end{center}
\begin{center}
\begin{figure*}
\mbox{
\includegraphics[scale=0.4,angle=90,width=8cm,height=10cm,keepaspectratio]{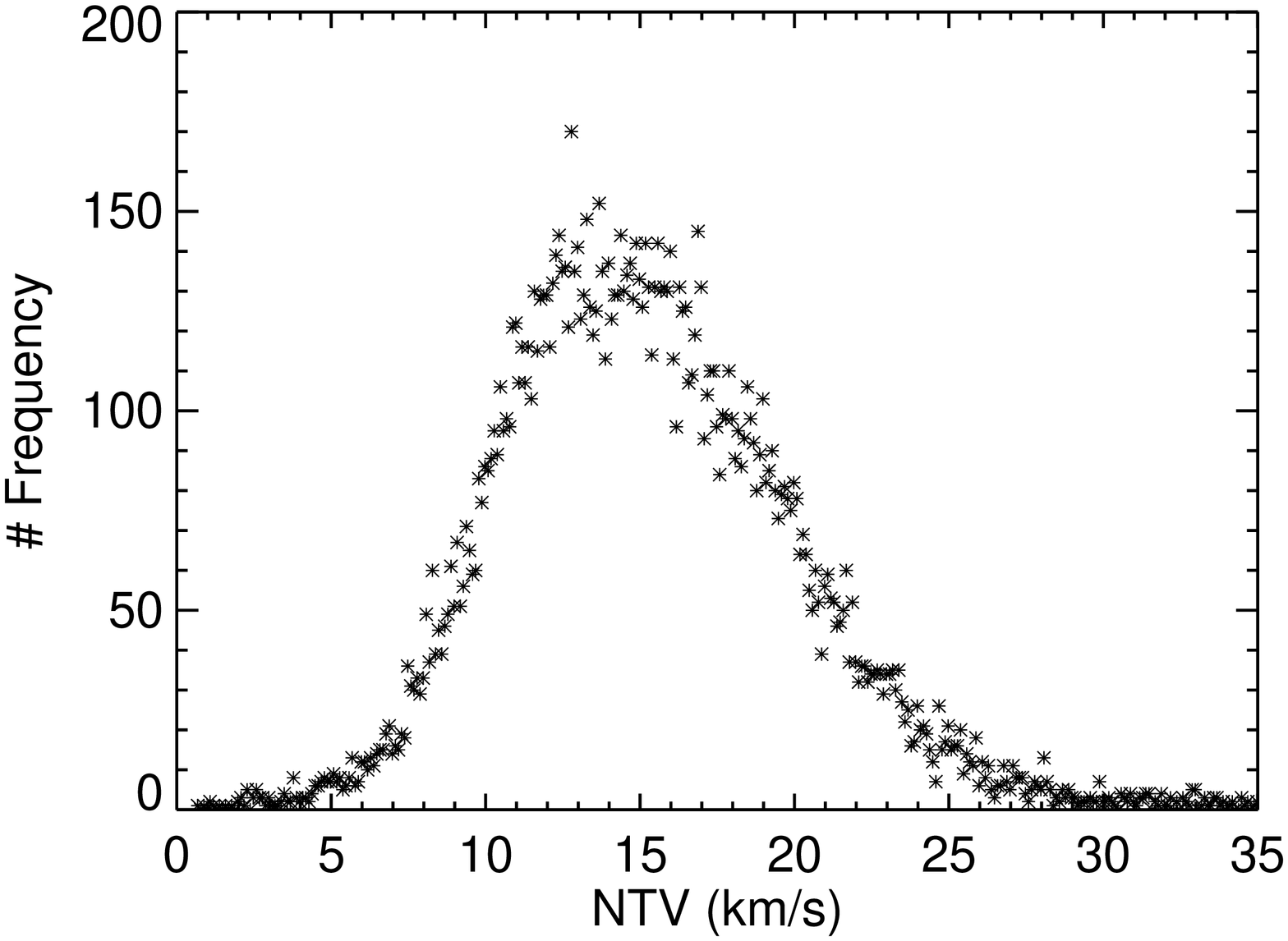}

\includegraphics[scale=0.4,angle=90,width=8cm,height=10cm,keepaspectratio]{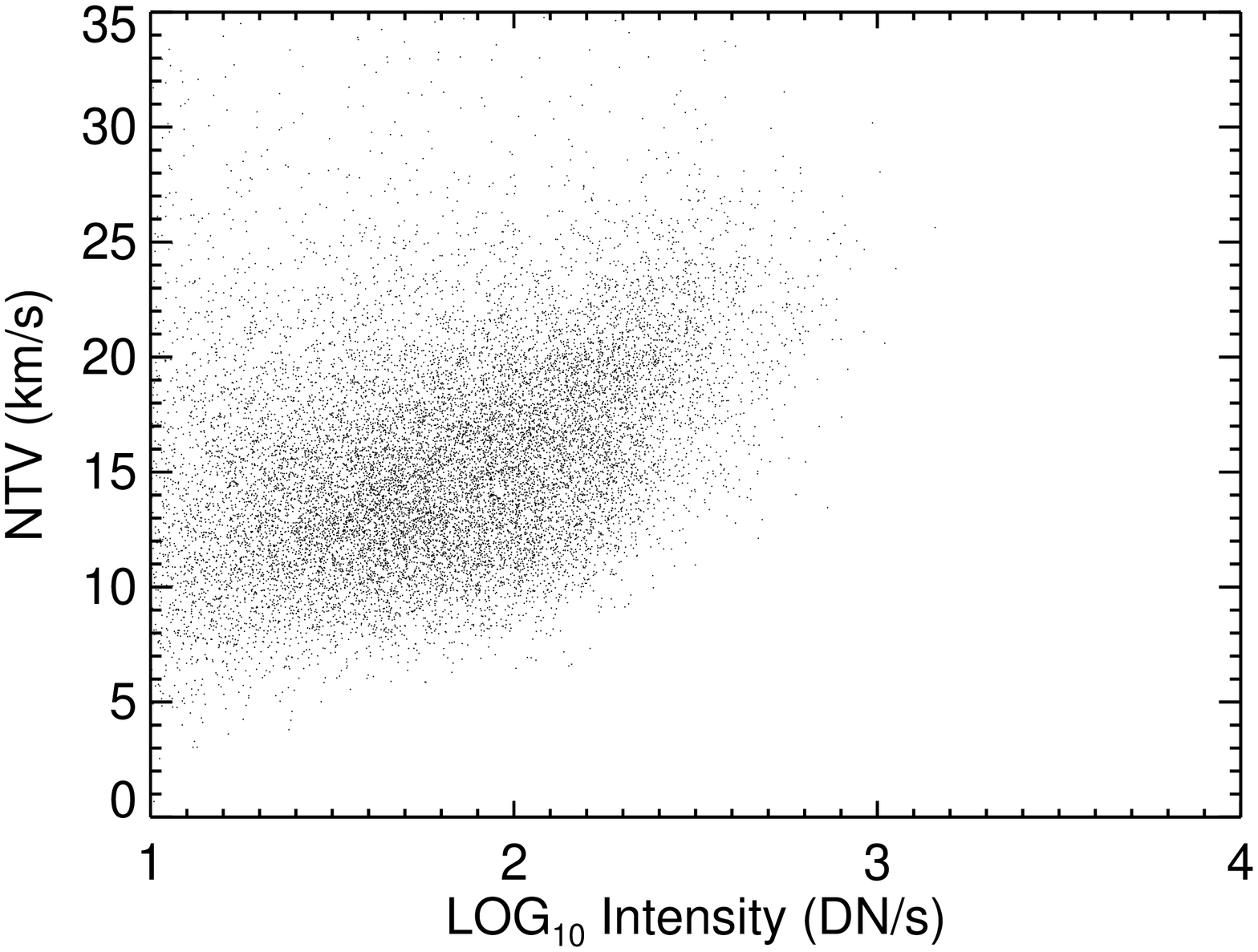}
}
\caption{Left panel: The distribution of non-thermal velocities for the for the QS observed on 25th Febraury 2014 for which the parametric plot is shown in Fig. \ref{3}. Right panel: Scatter plot of the NTVs as a function of the intensities of the \ion{Si}{IV} 1393.75 \AA\ spectral line.}
\label{4}
\end{figure*}
\end{center}
\section{Observational Data and Analyses}
We searched the IRIS database and selected
various spectral datasets where the IRIS slit rastered quiet Sun regions. 
IRIS provides spectral observations in the form of raster scans in the FUV domain (1332 – 1407 \AA) having various emission lines. We use the \ion{Si}{iv} line (1393.75 \AA) corresponding to the upper chromosphere/TR for our analysis. The observations have different exposure times ranging from 4 to 30~s. 
Most of the observations used in our work are 'Very Large' covering a region of 175\arcsec along the slit in the y-direction. The slit width is 0.33\arcsec. The whole raster scan covers different regions in the $x$-direction depending on the size of the raster step: dense - 0.33\arcsec; sparse - 1\arcsec; coarse - 2\arcsec.

We use well calibrated Level~2  data. All the technical effects such as dark current, flat fielding, and geometric correction are taken care of. Also, the spectral drifts caused by thermal variations and the spacecraft's orbital velocity are accounted for before analysis. In addition, we have removed residual cosmic rays using the solarsoft routine 
{\it new\char`_spike.pro}, although we note that spikes in the data are often present.
To increase the signal to noise ratio and have enough signal for the 
\ion{Si}{iv} lines, we re-binned the data to 1\arcsec\ in x and y-direction.
We recall that \citet{2015ApJ...799L..12D} showed that the resulting NTVs are very similar 
to those obtained at the full IRIS resolution.
We use a single Gaussian fitting and determine peak intensity, Doppler shift, and width of the lines using custom written software based on the {\it cfit} suite of programs developed for the analysis of SoHO CDS data. We also exclude the non-Gaussian profiles by imposing a chi-square condition that has values less than 5 (thus eliminating profiles which do not fit well to a Gaussian profile). Examples of spectral fits having averaged peak intensities around 400 DN and different values of chi-square are shown in Fig. \ref{spec_fit}. This shows that the chi-square values less than 5 are well-fitted with a single Gaussian and there are no obvious indications of broad wings. This constraint removes a fraction of about 20--30\% of the data. Further examples are provided in the Appendix. We find that this chi-square condition is reasonable also for weaker regions.

The instrumental broadening (FWHM) for IRIS is equivalent to 0.03 \AA. 
We assume that the 
\ion{Si}{iv} formation temperature is  
80 $\times$ 10$^{3}$ K which gives a 
thermal FWHM of 0.053\AA.
This is the temperature assumed by De Pontieu et al., and is obtained using the zero-density ionization equilibrium in CHIANTI (\citealt{2019ApJS..241...22D}). 
However, \citet{2021MNRAS.505.3968D} recently reported new ionization equilibrium calculations which include density-dependent effects and charge transfer, showing that the formation temperature of the \ion{Si}{iv} doublet is much lower. Taking into account the quiet Sun emission measure distribution, the \ion{Si}{iv} lines are predicted to be mainly formed around 60 $\times$ 10$^{3}$ K. However, we note that changing the formation temperature of \ion{Si}{iv}  by 2 $\times$ 10$^{4}$ K has little effect ($~$5\%) in the NTV measurements.

The details of all the observations are given in Table \ref{table1}. 
The first column shows the date and time of the observations. Their corresponding locations are also given. The first two observations (20th September 2013 and 4th October 2013) focus on the quiet-sun region near the limb with exposure times of 4 and 30~s, respectively. The third dataset of QS at the North Pole limb having exposure time of 15~s is featured in Fig. \ref{fig1}, \ref{fig2}. The last two observations on 25th February 2014 are of particularly interest as they are consecutive observations targeting the same quiet-Sun region but having different exposure times (8 and 30~s). The one having a 30~s exposure time is shown in the paper (Figures~\ref{3}, \ref{4}). The other observation is discussed in the Appendix along with all other observations analysed. In the main part of the paper, we show single observation from the limb (22nd January 2014) and disc centre (25th February 2014) to focus on the centre-to-limb variation.
The last column (Table \ref{table1}) shows the radial distance of the centre of the raster from the Sun's centre. The disc observations are close to the Sun's centre, having radial distances less than 0.5 R$_{\odot}$. These are compared with limb observations.
\subsection{Observational Results}
The fitting of the 1394 \AA\ line for the Quiet-Sun observed near the limb on 22nd January 2014, with an exposure time of 15~s, gives different parametric values (Intensity, Doppler velocity, Non-thermal velocity) at each pixel of the region which is shown in Fig. \ref{fig1}. The left panel shows the intensity, where the bright network regions are distinct from
the background region. The corresponding Doppler velocity and non-thermal widths are shown in the middle and right panels. The variation of NTVs with radial distance from sun's centre in Fig. \ref{fig1extra} shows an increase in the NTV towards the limb. It also shows a further increase in off limb locations, in agreement with the earlier observations.
Fig. \ref{fig2} shows the distribution of non-thermal velocities 
just from the selected FOV, excluding the limb and above the limb region peaking at around 20 km s$^{-1}$. The intensity is well correlated with the NTVs as shown in the right panel.  A Gaussian fit for the NTV distribution shows that the NTV peaks at 
18.6 km s$^{-1}$ with a width of 9.6 km s$^{-1}$.

A similar analysis was conducted for the Quiet-Sun region observed near the disc center having an exposure time of 30~s. The parametric plots are shown in Fig. \ref{3}. In this case, the NTV distribution peaks at around 15 km s$^{-1}$. Its correlation with intensity is shown in Fig. \ref{4}. 
A Gaussian fit for the NTV distribution shown in Fig. \ref{4} shows that the NTV peaks at 14.9 km s$^{-1}$ having a width of 10.4 km s$^{-1}$.

We then repeated the analysis for various datasets targeting the QS close to the limb as well as at the disc center, having exposure times varying from 4 to 30~s. We have fitted the NTV distributions with  Gaussians. The peak NTVs along with the widths (FWHM) and the averages of the NTV values for all these
datasets are shown in Table \ref{table1}. The distribution of the NTVs for all the observations show that the NTVs are consistently higher towards the limb (about 20 km s$^{-1}$) than near the disc center (15 km s$^{-1}$). All the details for these observations, exposure times, the distance range of the FOV of the region used for the NTV calculations are listed in the Table \ref{table1}. The plots of all the observations are shown in the Appendix.
This centre to limb variation is found to be  independent of different exposure times. 

\subsection{Possible Opacity Effect}

Spectral lines affected by opacity tend to have broader line profiles. 
The two \ion{Si}{iv} lines form a doublet and share a common ground level. The \ion{Si}{iv} 1393.757 \AA\ line has an atomic transition of 3s$^{2}$S$_{1/2}$ - 3p$^{2}$P$_{3/2}$ and \ion{Si}{IV} 1402.772 \AA\ of 3s$^{2}$S$_{1/2}$ -  3p$^{2}$P$_{1/2}$. Assuming that these lines are optically thin, their ratio should be 2, which is equal to the ratio of the oscillator strengths for the two lines. Ratios lower than 2 indicate opacity effects, whilst 
ratios above 2 could be due to resonant scattering, as found by 
\citet{2018A&A...619A..64G} for 2 \% of individual profiles in an active region.

\begin{center}
\begin{figure*}
\mbox{
\includegraphics[scale=0.4,angle=90,width=8cm,height=10cm,keepaspectratio]{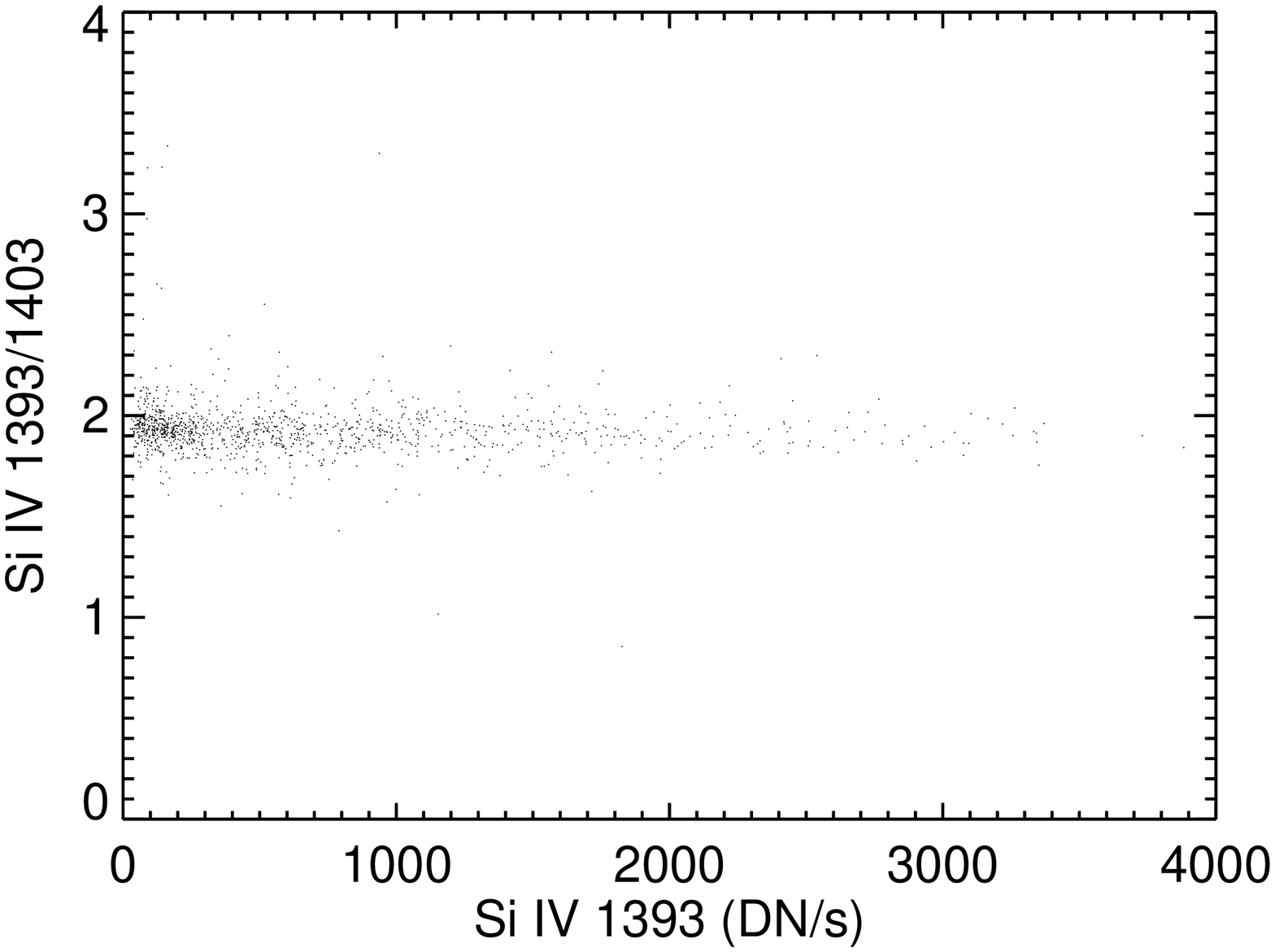}

\includegraphics[scale=0.4,angle=90,width=8cm,height=10cm,keepaspectratio]{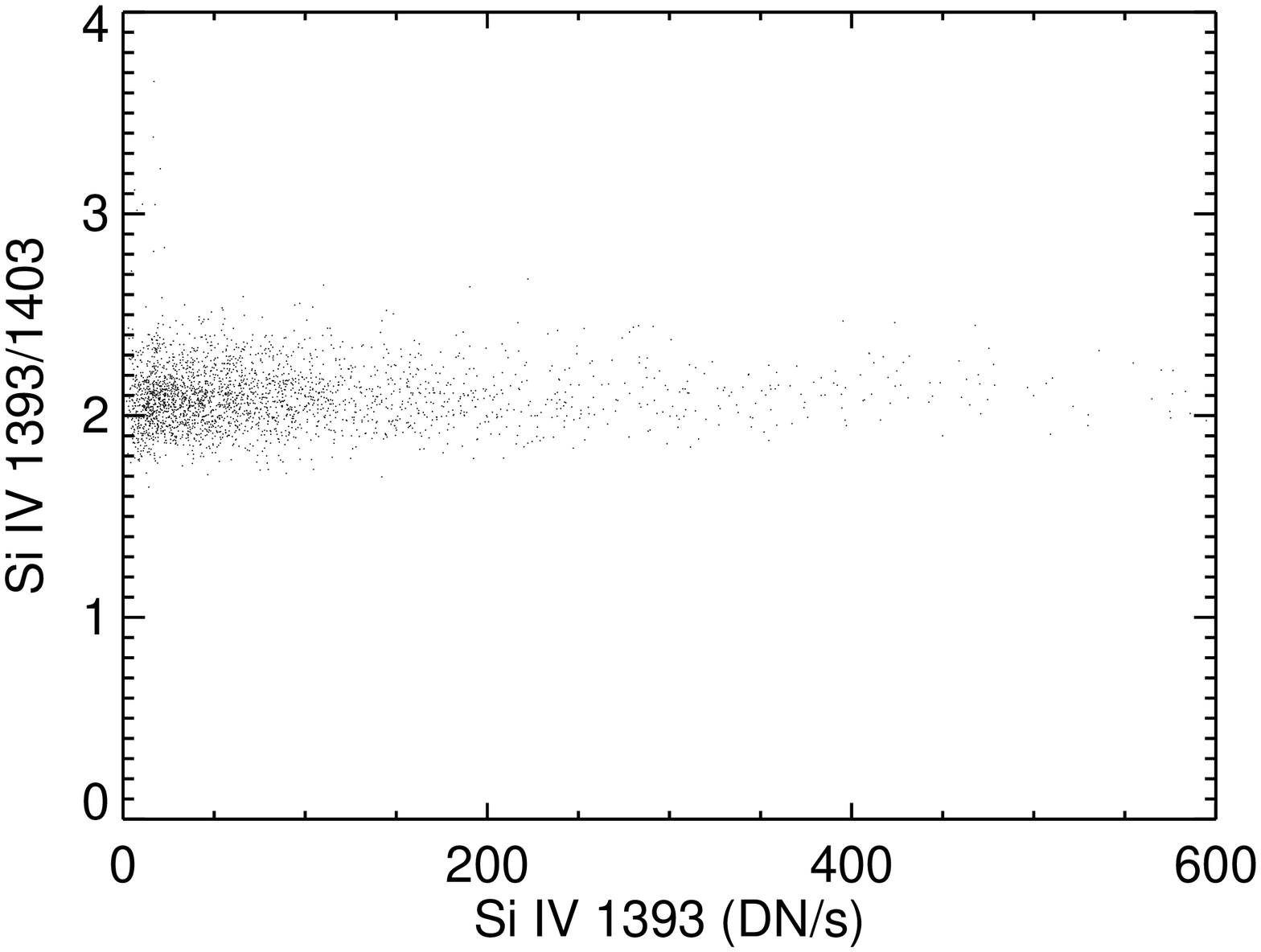}
}
\caption{Scatter plot of the line ratio \ion{Si}{iv} 1393/1403 \AA\ as a function of \ion{Si}{iv} 1393.75 \AA\ intensities for observations near the limb and disc center} in the left and right panel, respectively.
\label{5}
\end{figure*}
\end{center}

We generally observe the ratio to be around 2 in the datasets we analysed, as shown in Fig. \ref{5}. Similar results (i.e. no significant opacity estimated from line ratios) were found by previous authors in on-disc observations.  The distribution of the ratio of the two \ion{Si}{iv} (1394/1403 \AA\ ) lines are shown in 
Fig.~\ref{5} (left panel) for observations closer to the limb and
in Fig.~\ref{5} (right panel)  for observations near the disc center.
These are plotted for all the locations having chi-square value less than 5 and peak intensities greater than 200 DN, assuming all other locations to be noisy. These are the same criteria used for selecting the data used to calculate the NTVs in our paper. Ratios of 2 indicate that opacity has no strong effect with regard to the centre to limb variation of NTVs. 

\begin{table*}
\centering
\begin{tabular}{c c c c c c c}
\hline
Observation  & Raster   & Exposure   & Location   & Peak;Average of             &  Width & Distance Range \\
             &          &  Time(s)   &            & NTV (km s$^{-1}$)   &  (km s$^{-1}$) & (R$_{\odot}$) \\
 \hline
20/09/2013 (14:31) 		& Very Large Dense  & 4  & North Pole Limb & 18.4; 21.2 & 5.8  & 0.82 - 0.99 \\
04/10/2013 (15:01) 		& Very Large Dense  & 30 & East limb  & 19.2; 20.2 & 7.5 & 0.84 - 0.89 \\
22/01/2014 (01:48)      & Dense Synoptic    & 15 & North Pole Limb & 18.3; 19.63 & 6.4 & 0.83 - 0.99 \\
26/09/2013 (16:50\_r0)  & Very Large Sparse & 8 &  Disc  & 14.9; 17.0   & 8.0 & 0.38 - 0.57 \\
26/09/2013 (16:50\_r1)  & Very Large Sparse & 8 & Disc & 14.7; 16.5 & 8.6 & 0.38 - 0.57 \\
22/10/2013 (11:30)      & Very Large Dense & 30 &  Disc & 16.1; 16.4  & 7.3 & 0.25 - 0.44 \\
27/10/2013 (01:22)      & Very Large Coarse &  4 &  Disc & 16.5; 17.1  & 7.9 & 0.001 - 0.12 \\
25/02/2014 (18:59)      & Very Large Dense & 8 &  Disc & 15.3; 17.1  & 7.8 & 0.02 - 0.24 \\
25/02/2014 (20:50)      & Very Large Dense & 30 & Disc & 14.9; 15.4 & 6.8 & 0.004 - 0.23	\\
\end{tabular}
\caption{Non-thermal velocities observed for different IRIS observations at different locations targeting the QS, with different exposure times.}
\label{table1}
\end{table*}
\section{Discussion and Conclusions}

In this paper, we have studied the non-thermal velocities for \ion{Si}{iv} and their center to limb variation in the quiet Sun. Various IRIS observational datasets at different locations with different exposure times suggest that varying the temporal resolution has no effect on the NTVs. The datasets closer to the limb have NTVs that peak around 20 km s$^{-1}$ while the disc datasets have NTVs with peaks varying from 14-17 km s$^{-1}$. 
We also note that the distribution of NTVs is wider for the disc observations, with many values lower than 10 km s$^{-1}$. In contrast, there are fewer locations having NTV values less than 10 km s$^{-1}$ near the limb. In all cases, there is a clear correlation between intensities and NTVs, as found previously. 

We are not aware of any previous studies where the effect of the exposure time
was considered, although in some cases it was noted that long exposures could result in larger NTVs. This means that on average the timescales of the Doppler motions producing the NTVs are either relatively short, of the order of the 
exposure times considered here, or are irrelevant.

\cite{2019ApJ...886...46G} studied the  \ion{Si}{iv} Doppler velocities in an AR, finding weak center to limb variations.
They proposed an interpretation which included the effects of spicules. \cite{2016ApJ...827...99T} studied the \ion{Fe}{xii} coronal emission with IRIS, and compared two datasets, at disc center and closer to limb. They reached the opposite conclusions to our paper, i.e., that the \ion{Fe}{xii} data suggested more field aligned flows. This is most likely due to the \ion{Fe}{xii} originating only in a subset of the features emitting \ion{Si}{iv}, most of the  \ion{Fe}{xii} emission being connected to coronal loops.


\citet{1999ApJ...516..490P} reported a small decrease  (2$-$3  km s$^{-1}$) of the NTV towards the limb, while \citet{2000A&A...356..335D} and \cite{2008ApJ...673L.219M} reported increases of similar amounts above 0.9 R$_{\odot}$. Our results confirm these latter findings, with variations already present above 0.8 R$_{\odot}$. In addition, we have shown that the variations in the peak (or average) are also associated with variations in the widths of the distributions. We have also observed significant increases in the NTV above the limb, as pointed out in all previous studies.

As the IRIS 
instrumental width is very small, the variations we observe must be real and 
not due to instrumental effects.

After the Skylab observations, it was generally thought that the NTVs were isotropic, with the exception of the off-limb regions. 
For example, \citet{1979A&A....73..361M} studied optically thin emission from the QS using the NRL slit spectrograph, S082B, on Skylab. They reported that the limb broadening measurements were consistent with isotropic acoustic flux propagating through the transition region, also consistent with the disc broadening measurements in their earlier work (\citealt{1978ApJ...226..698M}).  

\cite{1998A&A...337..287E} used SUMER observation to claim that the NTVs were not isotropic, although inspection of their results shows that NTVs on-disc were not varying. Larger NTVs were only observed off-limb, as in the Skylab observations and in the present IRIS results.
\citet{1999ApJ...516..490P} found nearly constant NTVs in \ion{C}{iv}, with a small decrease towards the limb. \citet{2000A&A...356..335D} analysed the same observations but found the opposite behaviour. 
In any case, it is puzzling that a centre-to-limb variation is observed in \ion{Si}{iv} by IRIS but was not observed by SUMER in \ion{C}{iv}, also considering that these ions are formed at similar temperatures.
Both doublets in \ion{Si}{iv} and \ion{C}{iv} should be largely free of significant broadening due to opacity,
as the ratios are always close to 2 in the quiet Sun ( \citealt{1993SoPh..144..217D}). 

Most of the Doppler motions in \ion{Si}{iv} cluster
around 15 km s$^{-1}$, a value significantly lower than that found in most
previous literature, except the earlier results from Boland et al., 
those from the Spacelab-2 HRTS dataset.
It is interesting to note that all the SUMER results with larger NTV were obtained with much longer exposure times, so it is clear that long exposures ($>$ 100 s) tend to produce higher NTVs. It is also likely that the larger NTVs reported in the literature are affected by including non-Gaussian profiles, most of which occur during explosive events in the supergranular network.

If the very small spectroscopic filling factors in the 
transition region \citep[see, e.g.][]{dere_etal:1987} are interpreted as real volume filling factors \citep[see also][]{Judge_2000}, the subresolution structures would have sizes of the order of 3 to 30 km, hence would be unresolved 
at our binned IRIS resolution (1 \arcsec), but also at the native IRIS resolution.
As a consequence,  the observed NTVs would be the effect of a superposition of different flows along the line of sight, occurring on such small spatial scales and perhaps
with short durations.

We do not currently have instruments which could observe such flows in TR emission, but we can obtain some information from observations in the chromosphere, where shorter exposure times and higher spatial resolutions have been achieved.
In fact, there is a close temporal and spatial connection between 
chromospheric and TR features. This was already noted from e.g. Skylab
and HRTS observations (see e.g. \citealt{1986ApJ...305..947D}).
Observations of cool material injected into the corona,
such as that in threads of prominences, show emission in transition region lines that is
co-spatial and co-temporal with lower-T chromospheric emission (\citealt{1979A&A....73..361M}).
IRIS observations have also confirmed this. 

There are many high-resolution, high-cadence chromospheric observations 
of Doppler flows.
In the last decade, rapid blue-shifted excursions (RBEs) in the blue wings of chromospheric spectral lines such as the Hydrogen H-$\alpha$ (observed in absorption)  have been observed. RBEs have Doppler velocities in the range between 10 and 30 km s$^{-1}$ and lifetimes between about 5~s to 50~s (see e.g.  \citealt{2013ApJ...769...44S} and references therein), although shorter lifetimes cannot be ruled out as 
most observations have cadences of about 8~s or longer.
 
It has been suggested that RBEs are the disc counterpart of type II spicules (\citealt{2013ApJ...764..164S}), which are much shorter lived
(tens of seconds) than the type I spicules (minutes).
However, note that the apparent velocities of  type II spicules
are between 50 and 150~km.

Recently, rapid redshifted excursions (RREs) have also been observed in the
red wings of chromospheric lines.
Both RBEs and RREs tend to be present in the same locations and have similar lengths, widths, lifetimes and Doppler signatures (see e.g. \citet{2013ApJ...764..164S}).
Most of the Doppler velocities for both excursions range between 10 and 20 km s$^{-1}$ for the \ion{Ca}{ii} 8542~\AA\ and between 20 and 35 km s$^{-1}$ for the H$\alpha$. RBEs are more abundant than RREs. Oscillatory swaying motions in type II spicules are common and have amplitudes of the order of 10–20 km s$^{-1}$ and periodicities of 100–500 s 
(\citet{2007PASJ...59S.655D}).

Observations in the red wing of the H$\alpha$ line with a cadence of about 1~s showed the presence of
many fine structures over timescales of just a few seconds, and very high apparent velocities \citep{2012A&A...544A..88J}.
The authors suggested that some of the events could result from plasma sheet structures in the chromosphere.

So it is natural to expect that the RBEs and RREs seen in absorption in chromospheric lines would also
have a counterpart emission in TR lines such as \ion{Si}{iv}.
The Doppler velocities have been observed with a spatial resolution of a fraction of an arcsecond.
The Doppler velocities in the rapid excursions in the \ion{Ca}{ii} 8542~\AA\ line and the swaying motions of type II spicules have velocities within the range of values of the observed NTVs in \ion{Si}{iv}, so it is likely that these features are related. 

The fact that NTVs increase towards the limb and are even greater
off-limb, indicates that on average the Doppler motions are not aligned with the radial direction, and are mostly perpendicular. The NTVs could then be caused by swaying and torsional motions, which are commonly observed 
\citep[see, e.g.][]{2014Sci...346D.315D}. 
One would expect that the effects of the swaying and torsional motions would be enhanced in observations toward the limb, and decrease with short exposure times, which is what we observe.
Physically, Alfven wave heating would 
produce larger NTV near the limb, as shown e.g. by \cite{1998A&A...337..287E}.

\section{Acknowledgements}
 We would like to thank the anonymous referee for very useful comments. We would also like to acknowledge the financial support by STFC (UK) under the Research Grant ref: ST/T000481/1 and the research facilities provided by Department of Applied Mathematics and Theoretical Physics (DAMTP), University of Cambridge.
We acknowledge the use of IRIS observations. IRIS is a NASA small explorer mission developed and operated by Lockheed Martin Solar and Astrophysics Laboratory (LMSAL) with mission operations executed at NASA Ames Research Center and major contributions to downlink communications funded by the Norwegian Space Center (NSC, Norway) through an ESA PRODEX contract. We also acknowledge the use of the CHIANTI database. CHIANTI is a collaborative project involving George Mason University, the University of Michigan, the NASA Goddard Space Flight Centre (USA) and the University of Cambridge (UK).

\section{Data Availability}
All the data used for the analysis in our paper is available at IRIS website hosted by Lockheed Martin Solar and Astrophysics Laboratory (LMSAL): https://iris.lmsal.com/search/

\bibliographystyle{mnras}   	
\bibliography{mybib}  		
			     	
\section{Appendix}
\setcounter{figure}{0} 
\renewcommand\thefigure{A\arabic{figure}}
We have carried out an analysis of several more datasets to observe the centre to limb variation of non-thermal velocities. Fig. \ref{1.1} shows the parametric maps for the QS observed on the 20th September, 2013 close to the North limb having an exposure time of 4 s. Fig. \ref{1.2} shows the NTVs peak at 19.4 km/s with a correlation with intensity indicating that the bright regions have higher NTVs. Fig. \ref{1.15} shows the parametric maps for the QS region observed near an AR on the limb side on the 4th October, 2013 close to the East limb having an exposure time of 30 s. Fig. \ref{1.16} shows the NTVs peak at 19.3 km/s with a correlation with intensity indicating that the bright regions have higher NTVs.

Fig. \ref{1.3} shows the parametric maps for the QS observed on the 26th September, 2013 (Raster 0) near the disc center having an exposure time of 8 s. Fig. \ref{1.4} shows that the distribution of NTVs peaks at 14.9 km s$^{-1}$. 
Fig. \ref{1.5} shows the parametric maps for the QS observed on the 26th September, 2013 (Raster 1) near the disc center having an exposure time of 8 s. Fig. \ref{1.6} shows that the distribution of NTVs peaks at 14.7 km s$^{-1}$. 
Fig. \ref{1.7} shows the parametric maps for the QS observed on the 22nd October, 2013 near the disc center having an exposure time of 30 s. Fig. \ref{1.8} shows that the distribution of NTVs peaks at 16.1 km s$^{-1}$. Fig. \ref{1.9} shows the parametric maps for the QS observed on the 27th October, 2013 near the disc center having an exposure time of 4 s. Fig. \ref{1.10} shows that the distribution of NTVs peaks at 16.5 km s$^{-1}$.
Fig. \ref{1.13} shows the parametric maps for the QS observed on the 25th Febraury, 2014 near the disc center having an exposure time of 8 s. Fig. \ref{1.14} shows that the distribution of NTVs peaks at 15.3 km s$^{-1}$. All these observations are summarised in Table \ref{table1} along with other two observations discussed in the paper.

Finally, Figures~\ref{spec_fit1.2},\ref{spec_fit1.3},\ref{spec_fit1.4}
show a randomly-selected sample of spectral fits of IRIS 1\arcsec\ 
pixels, where the peak intensities of the lines range from 200 to 800 data numbers (DN).
The samples are ordered by their chi-square values. We recall that to measure the NTVs we selected spectral fits
with chi-square values less than 5, where the line profiles are nearly Gaussian.  

As the observations we analysed were taken at different times (hence the signal is affected by the instrument degradation) and with different exposures, there is a wide range of peak intensities in the 
spectra, ranging from about 200 to  800 DN (excluding bright explosive events). Plots of the chi-square values as a function of the peak intensities do not show any correlations, i.e. 
there is no tendency for the line profiles to become strongly non-Gaussians 
for weaker or larger intensities. In other words, we can find nearly Gaussian profiles of different widths for very low or very high intensities. 
There is a large scatter of values in the widths, as we have shown, but the tendency for having larger NTVs for higher intensities is clear in all observations.
Clearly, the line profiles in the lowest intensity regions with peak DN around 200 are more noisy than the other ones, but Gaussian fitting is still reliable. We recall that all the spectra were spatially averaged over 1\arcsec, to reduce noise. As most IRIS rasters were `dense', most of the averaging shows the spectra in a real 1\arcsec\ region of the quiet Sun. Fig. \ref{fwhm} shows the observed FWHM of two different \ion{Si}{iv} lines showing strong correlation between two lines

\begin{center}
\begin{figure*}
\includegraphics[scale=0.4,angle=90,width=15cm,height=12cm,keepaspectratio]{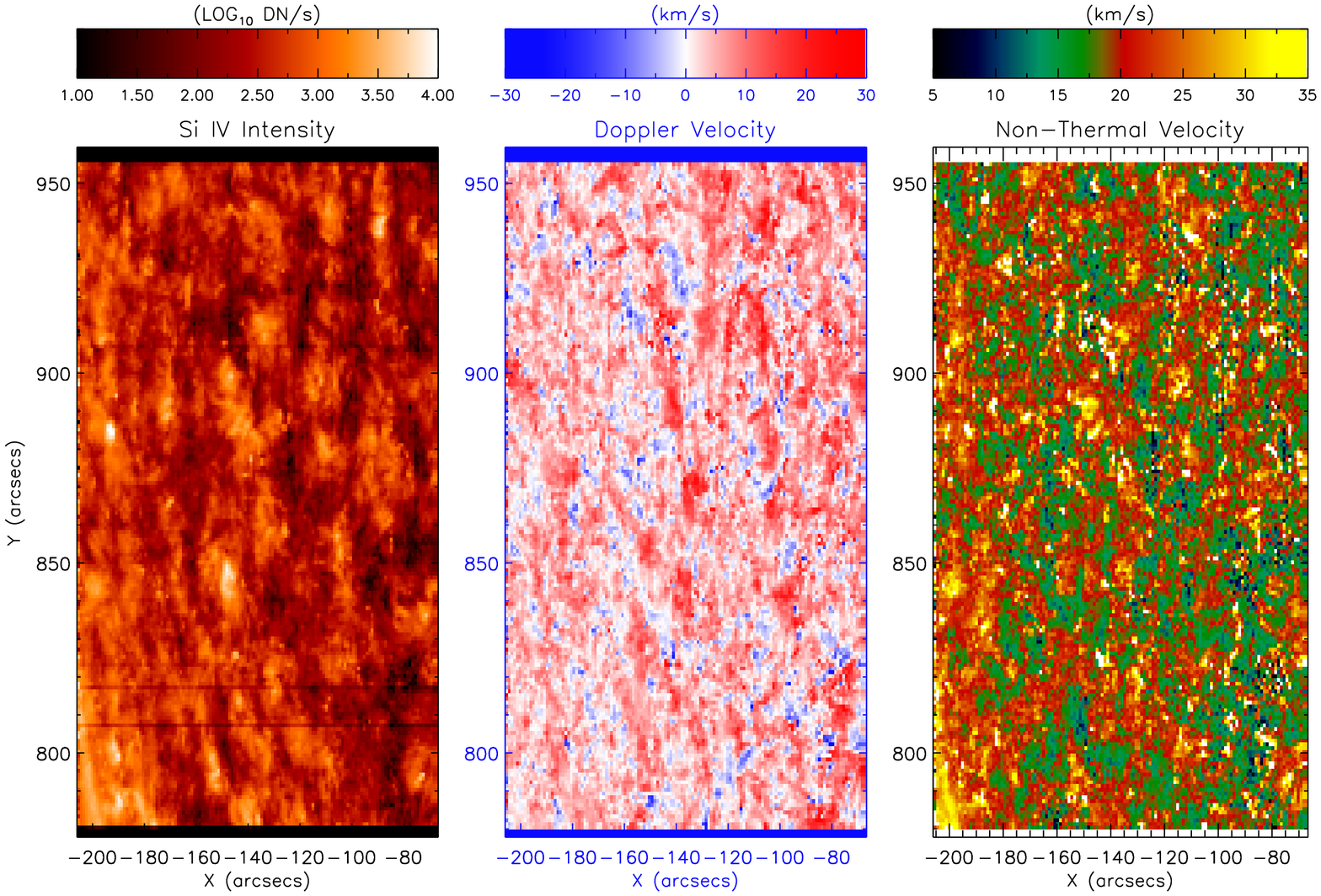}
\caption{Parametric plots of the QS observed on the 20th September, 2013 at the North limb having an exposure time of 4 s. The left panel shows the intensity of the IRIS \ion{Si}{iv} 1393.75 \AA\ line. The middle and right panels indicate the corresponding Doppler and Non-thermal velocities. The colour bars are shown above the plots.}
\label{1.1}
\end{figure*}
\end{center}

\begin{center}
\begin{figure*}
\mbox{
\includegraphics[scale=0.4,angle=90,width=8cm,height=10cm,keepaspectratio]{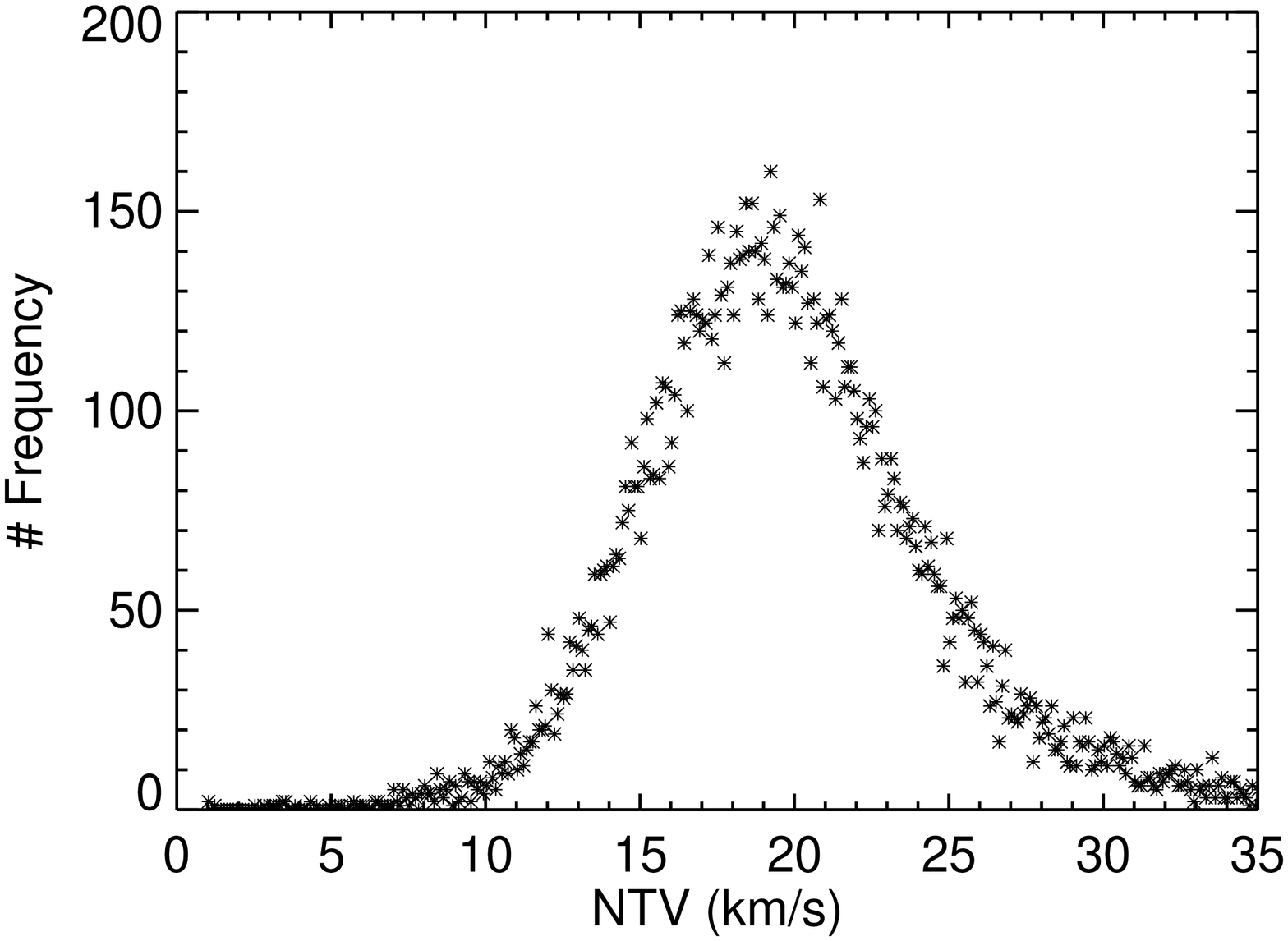}

\includegraphics[scale=0.4,angle=90,width=8cm,height=10cm,keepaspectratio]{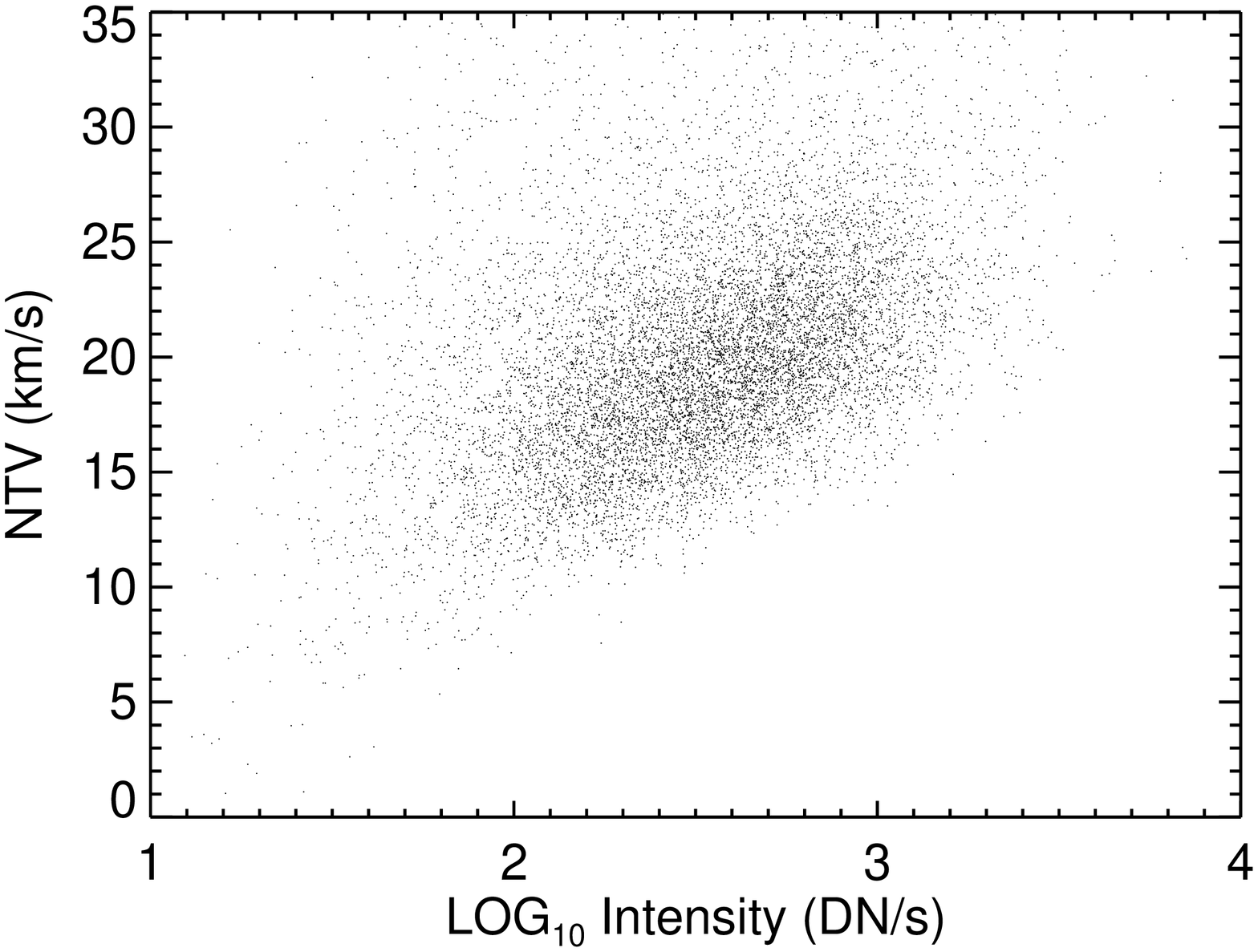}
}
\caption{Left panel: The distribution of non-thermal velocities for the QS observed on the 20th September, 2013 for which the parametric plot is shown in Fig.~\ref{1.1}. Right panel: Scatter plot of NTVs as a function of the intensity of the \ion{Si}{iv} 1393.75 \AA\ spectral line.}
\label{1.2}
\end{figure*}
\end{center}

\begin{center}
\begin{figure*}
\includegraphics[scale=0.4,angle=90,width=15cm,height=12cm,keepaspectratio]{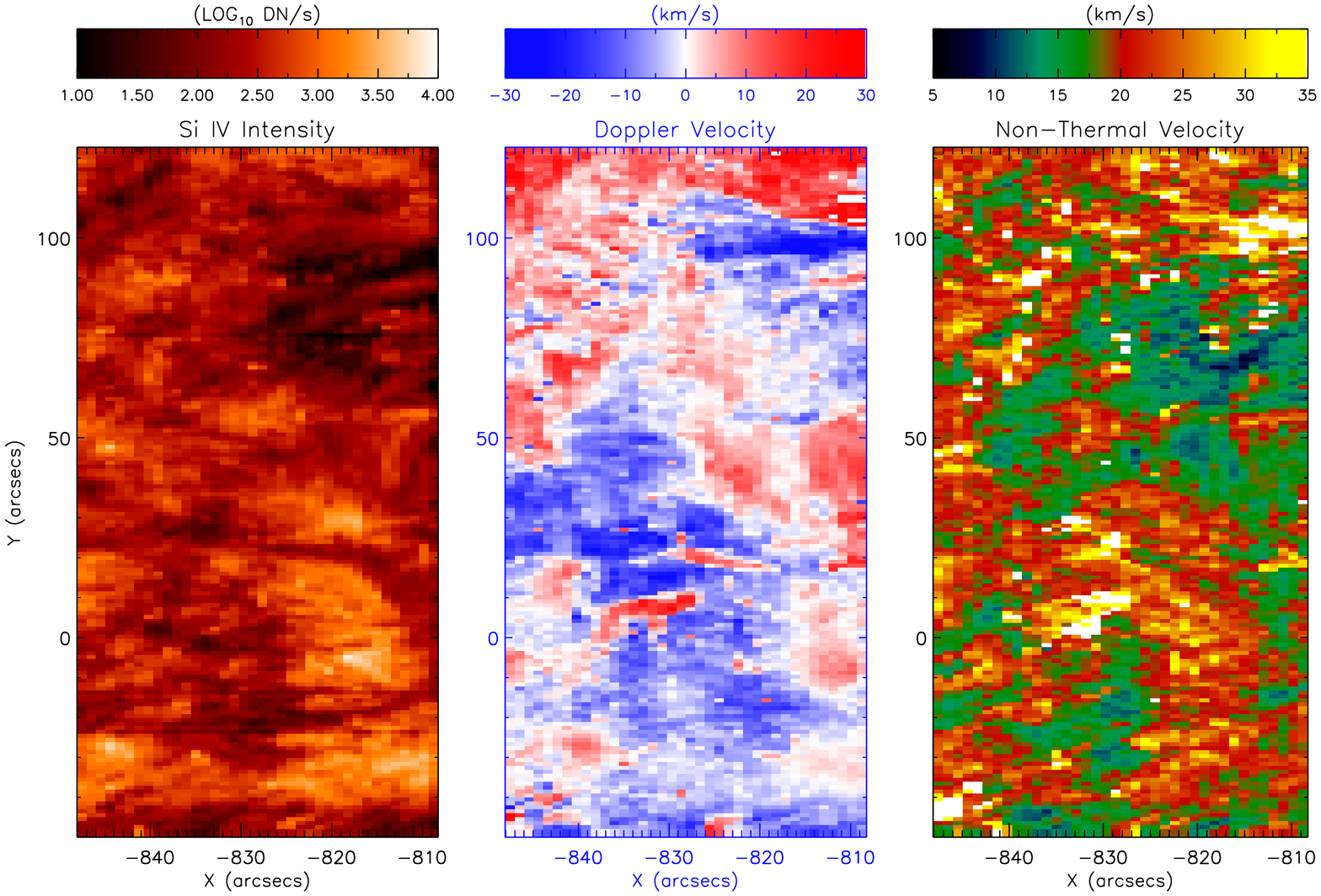}
\caption{Parametric plots of the QS observed on 4th October 2013, close to the East limb with an exposure time of 30 s. The left panel shows the intensity of the IRIS \ion{Si}{iv} 1393.75 \AA\ line. The middle and right panels indicate the corresponding Doppler and Non-thermal velocities. The colour bars are shown above the plots.}
\label{1.15}
\end{figure*}
\end{center}

\begin{center}
\begin{figure*}
\mbox{
\includegraphics[scale=0.4,angle=90,width=8cm,height=10cm,keepaspectratio]{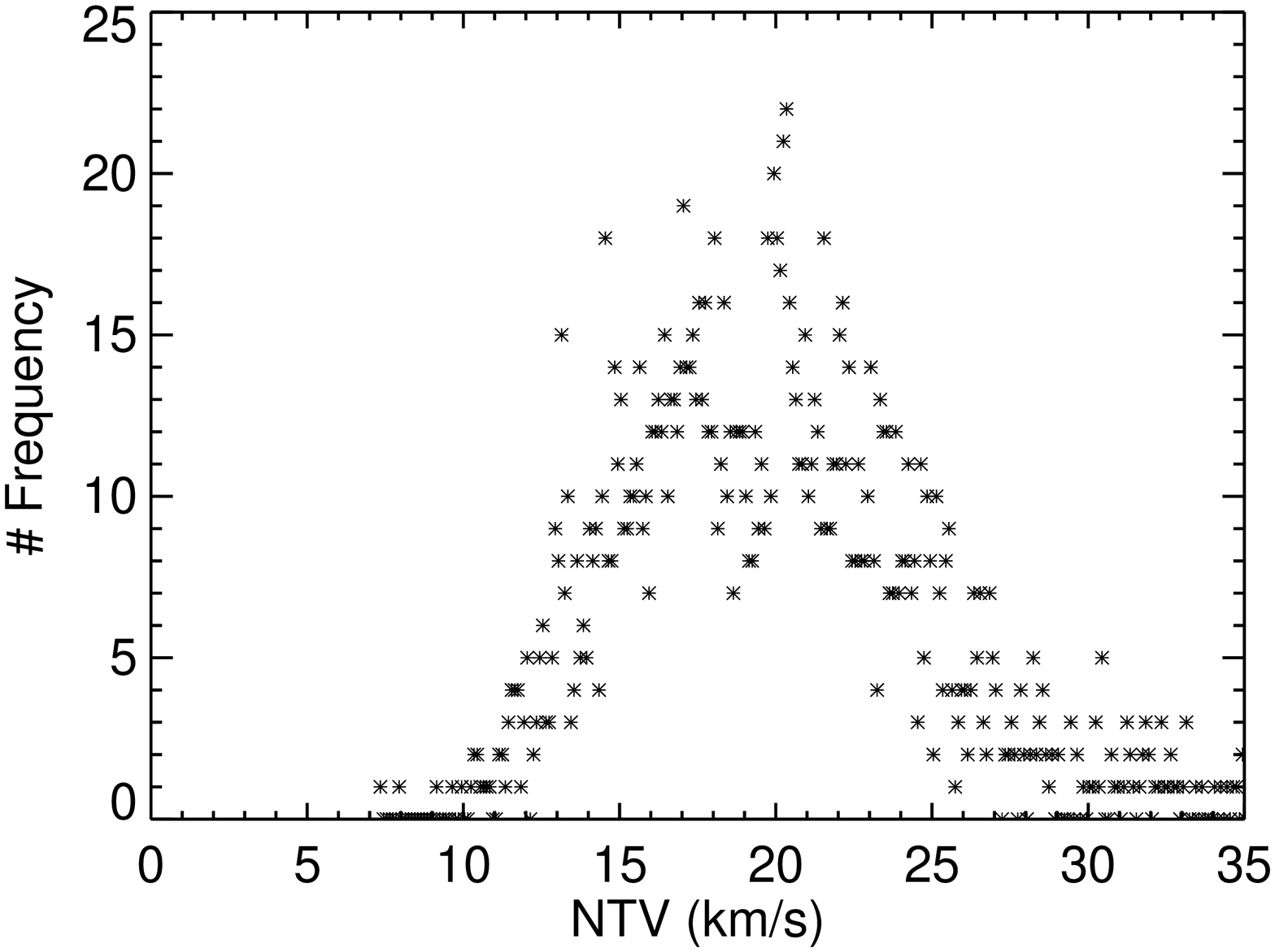}

\includegraphics[scale=0.4,angle=90,width=8cm,height=10cm,keepaspectratio]{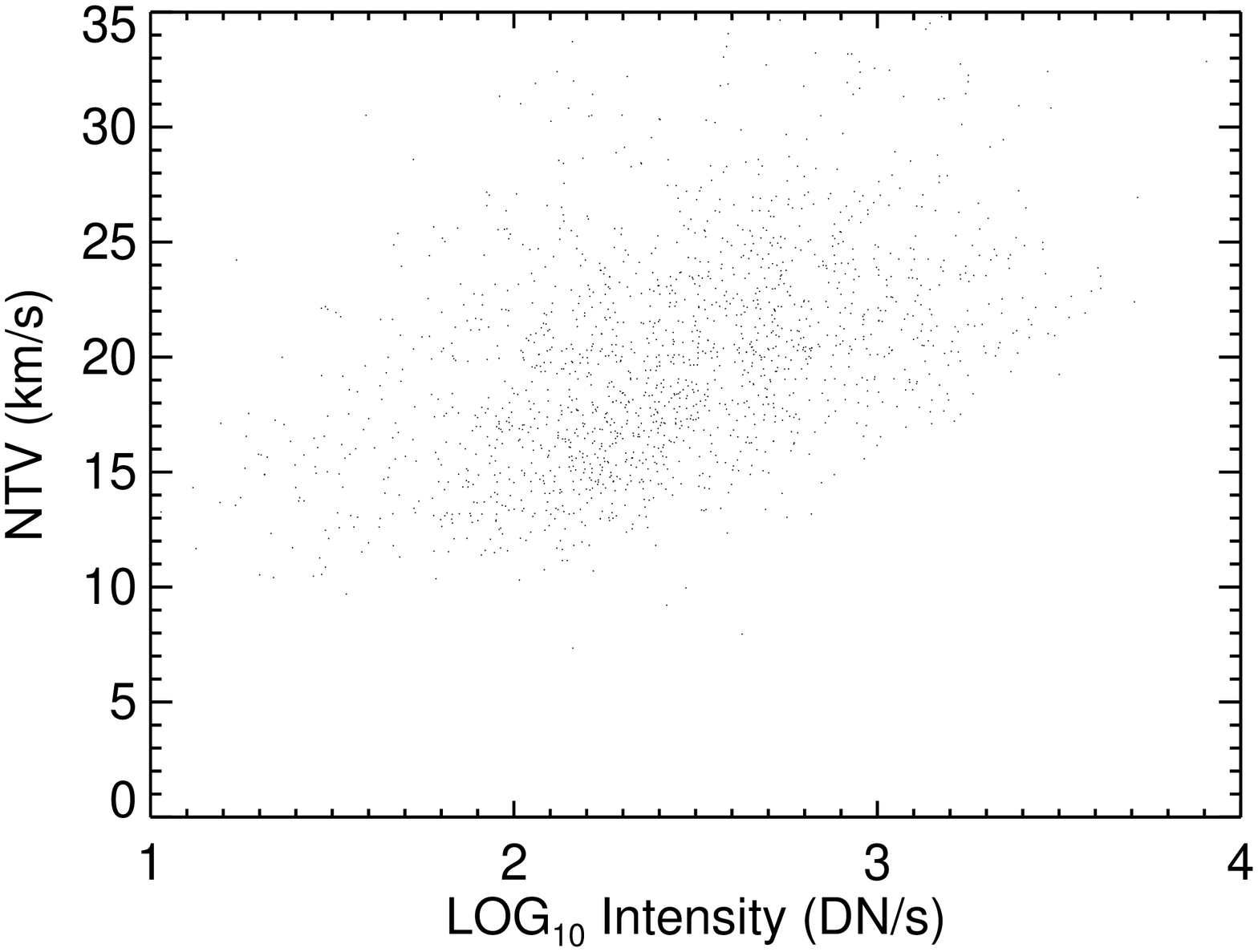}
}
\caption{Left panel: The distribution of non-thermal velocity for the QS observed on 4th October 2013 for which the parametric plot is shown in Fig.~\ref{1.15}. Right panel: A scatter plot of the NTVs as a function of the intensities of the \ion{Si}{iv} 1393.75 \AA\ spectral line.}
\label{1.16}
\end{figure*}
\end{center}

\begin{center}
\begin{figure*}
\includegraphics[scale=0.4,angle=90,width=15cm,height=12cm,keepaspectratio]{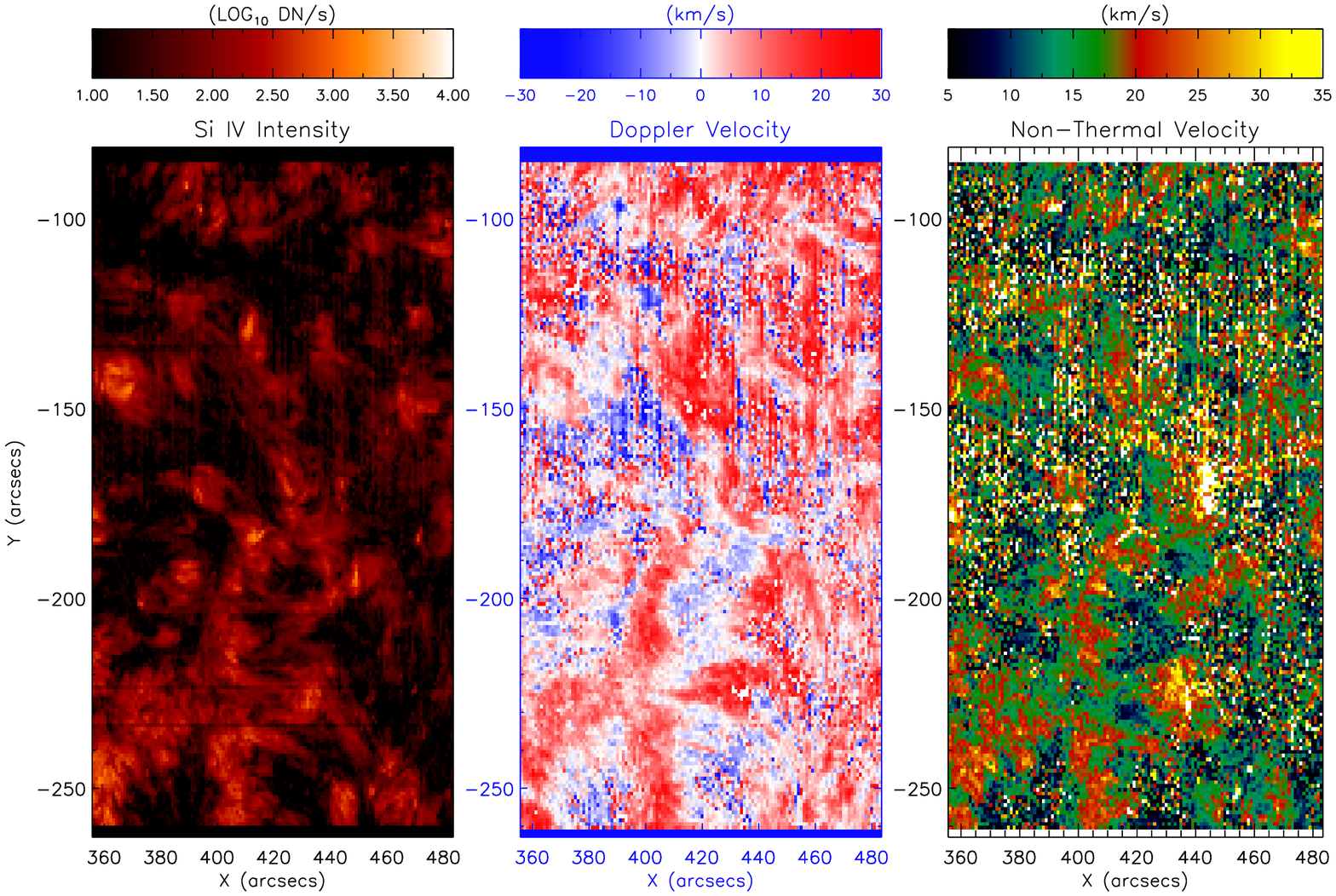}
\caption{Parametric plots of the QS observed on the 26th September, 2013 (Raster 0) near the disc center having an exposure time of 8 s. The left panel shows the intensity of the \ion{Si}{iv} 1393.75 \AA\ line. The middle and right panel indicates the corresponding Doppler velocities and non-thermal velocities. The colour bars are shown above the plots.}
\label{1.3}
\end{figure*}
\end{center}

\begin{center}
\begin{figure*}
\mbox{
\includegraphics[scale=0.4,angle=90,width=8cm,height=10cm,keepaspectratio]{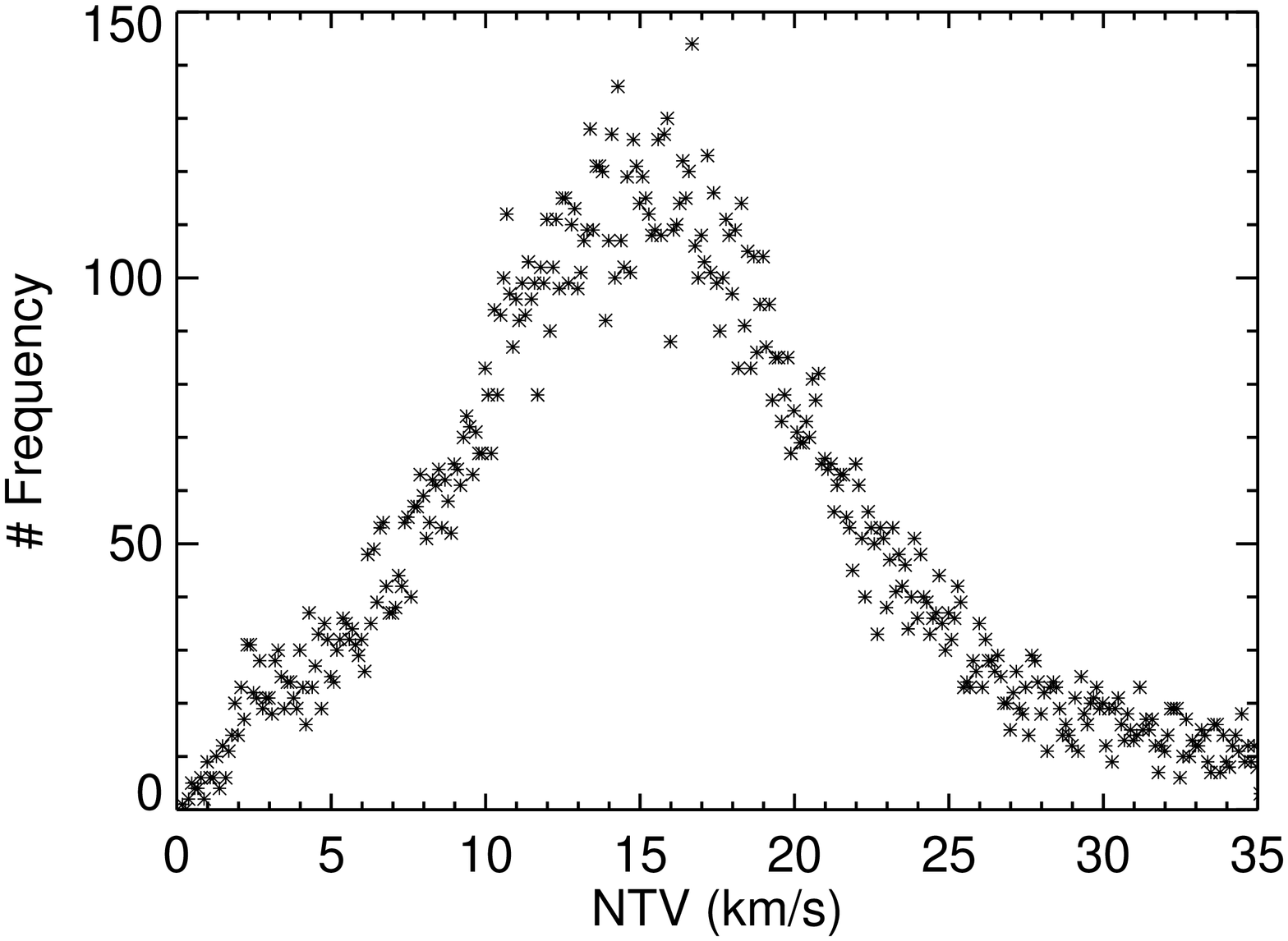}

\includegraphics[scale=0.4,angle=90,width=8cm,height=10cm,keepaspectratio]{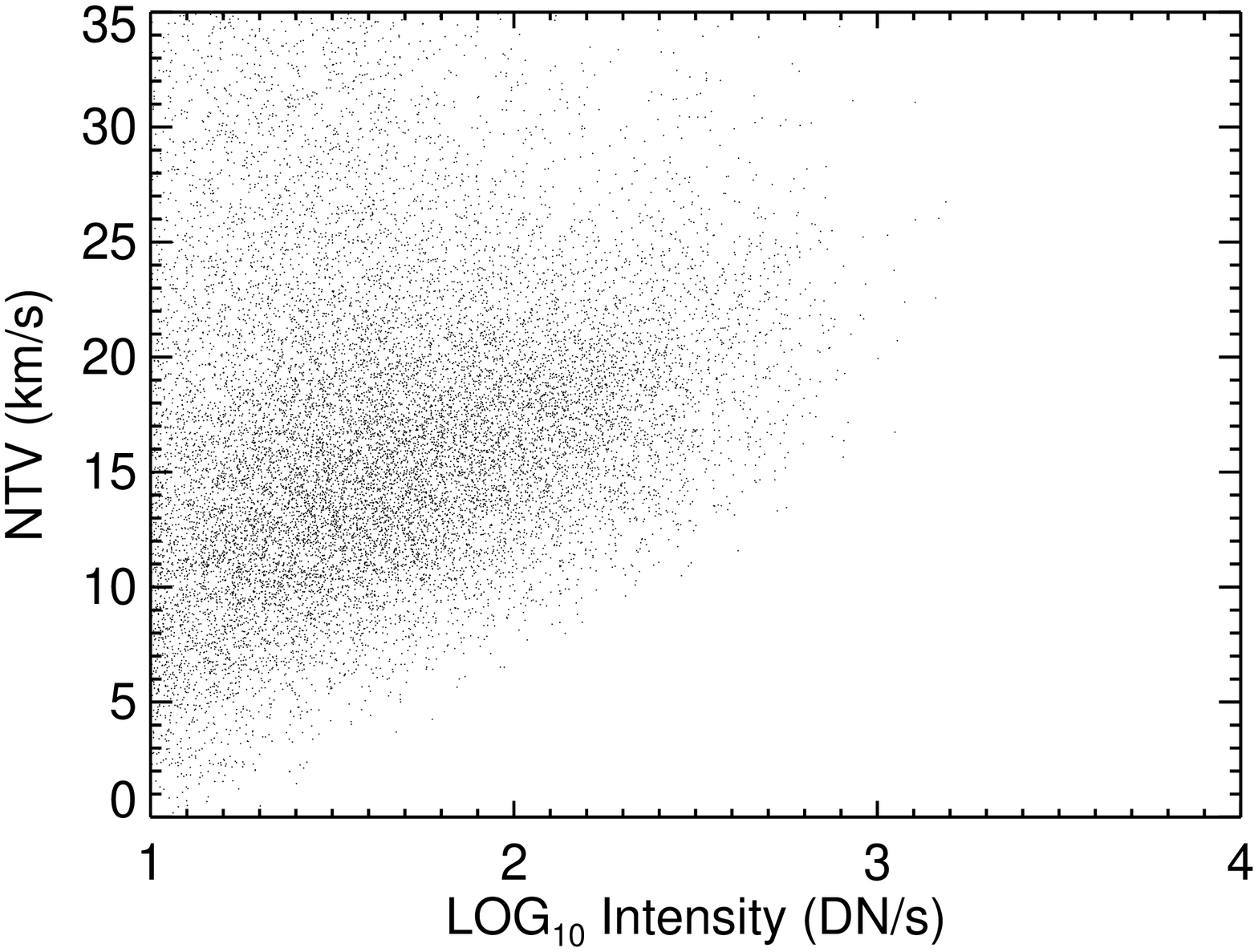}
}
\caption{Left panel: The distribution of non-thermal velocity for the QS observed on the 26th September, 2013 (Raster 0) for which parametric plot is shown in Fig.~\ref{1.3}. Right panel: Scatter plot of the NTVs as a function of the intensity  of the \ion{Si}{iv} 1393.75 \AA\ spectral line.}
\label{1.4}
\end{figure*}
\end{center}

\begin{center}
\begin{figure*}
\includegraphics[scale=0.4,angle=90,width=15cm,height=12cm,keepaspectratio]{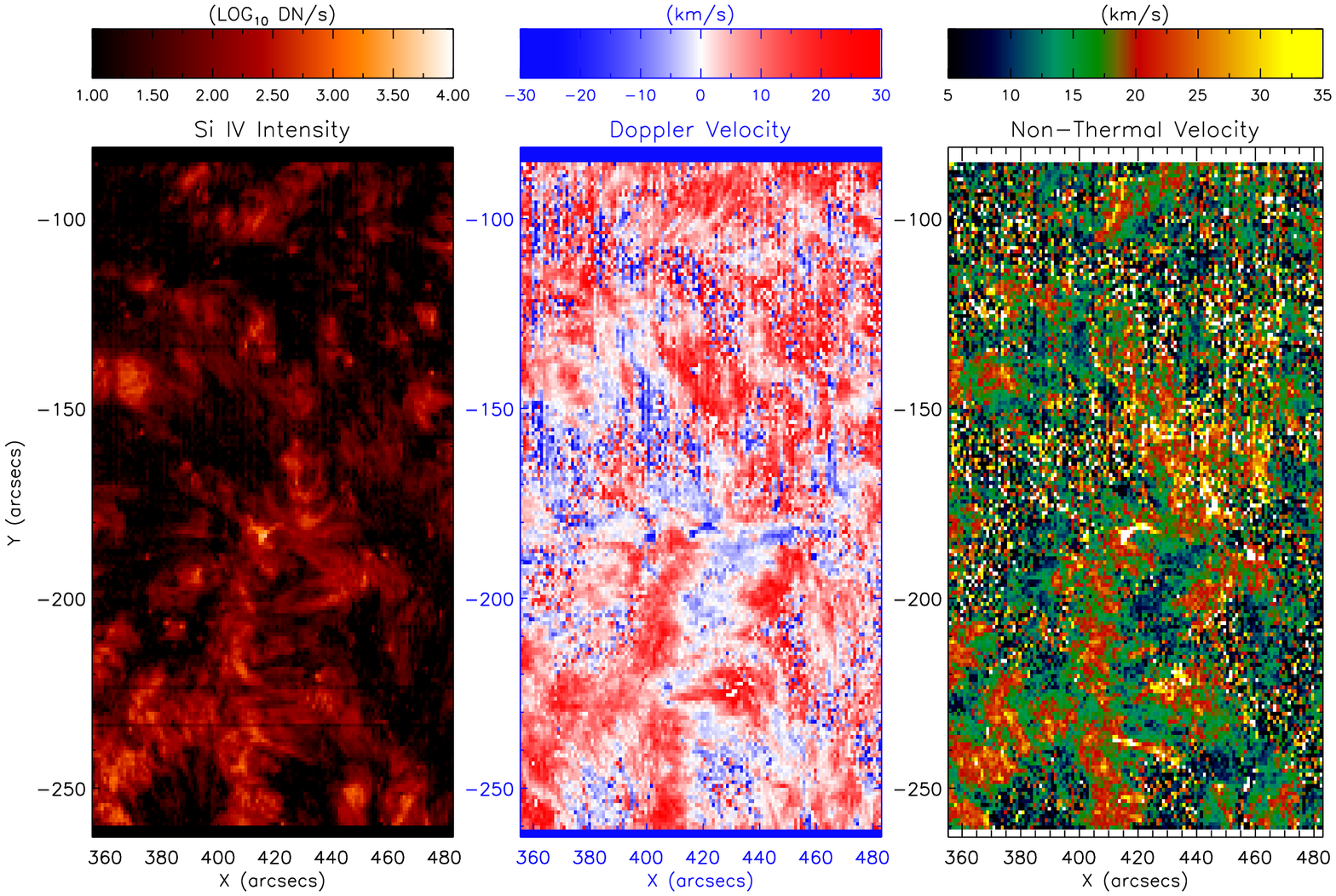}
\caption{Parametric plots of the QS observed on the 26th September, 2013 (Raster 1) near the disc center having an exposure time of 8 s. The left panel shows the intensity of the IRIS \ion{Si}{iv} 1393.75 \AA\ line. The middle and right panels indicate the corresponding Doppler and Non-thermal velocities. The colour bars are shown above the plots.}
\label{1.5}
\end{figure*}
\end{center}

\begin{center}
\begin{figure*}
\mbox{
\includegraphics[scale=0.4,angle=90,width=8cm,height=10cm,keepaspectratio]{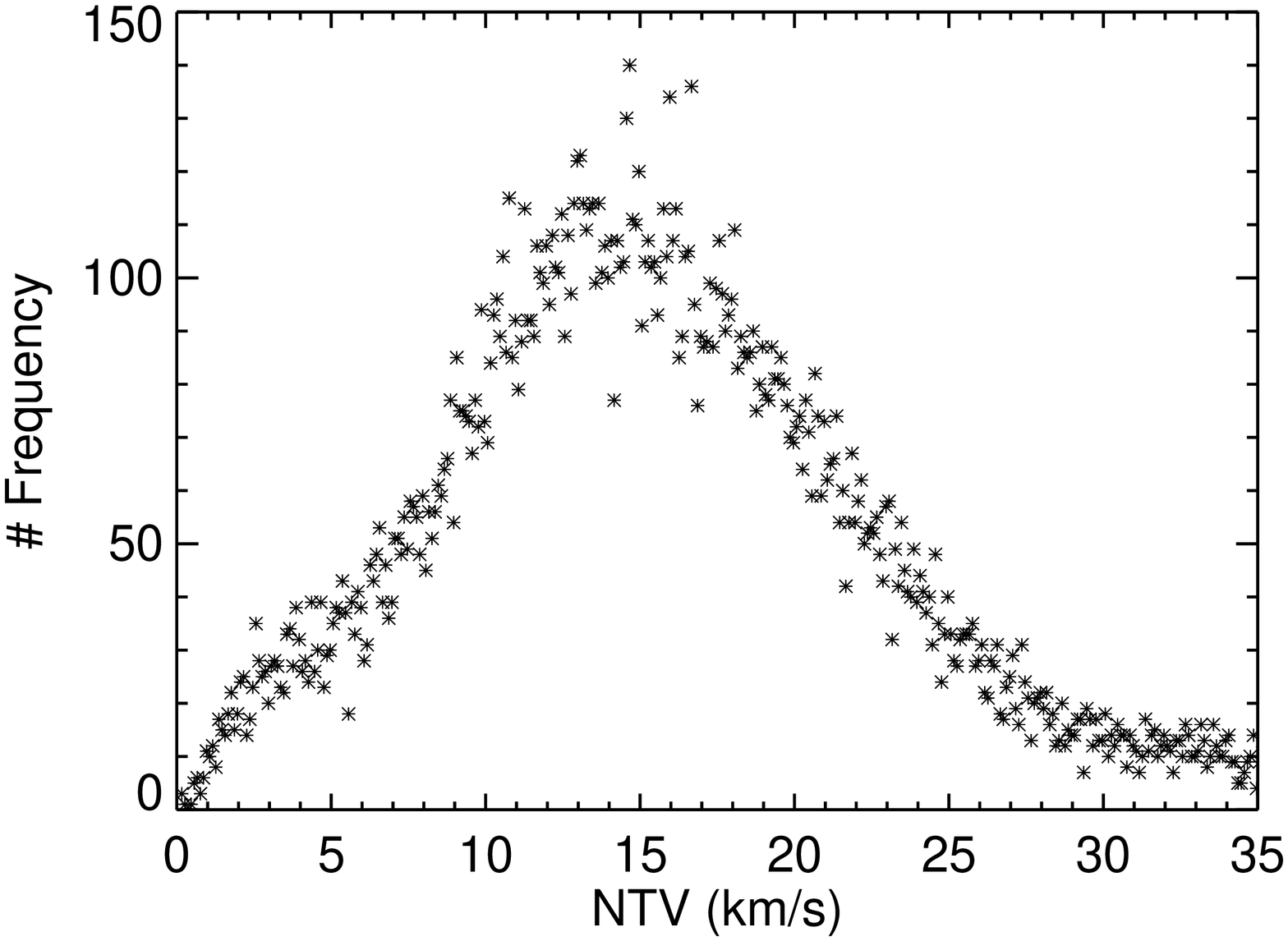}

\includegraphics[scale=0.4,angle=90,width=8cm,height=10cm,keepaspectratio]{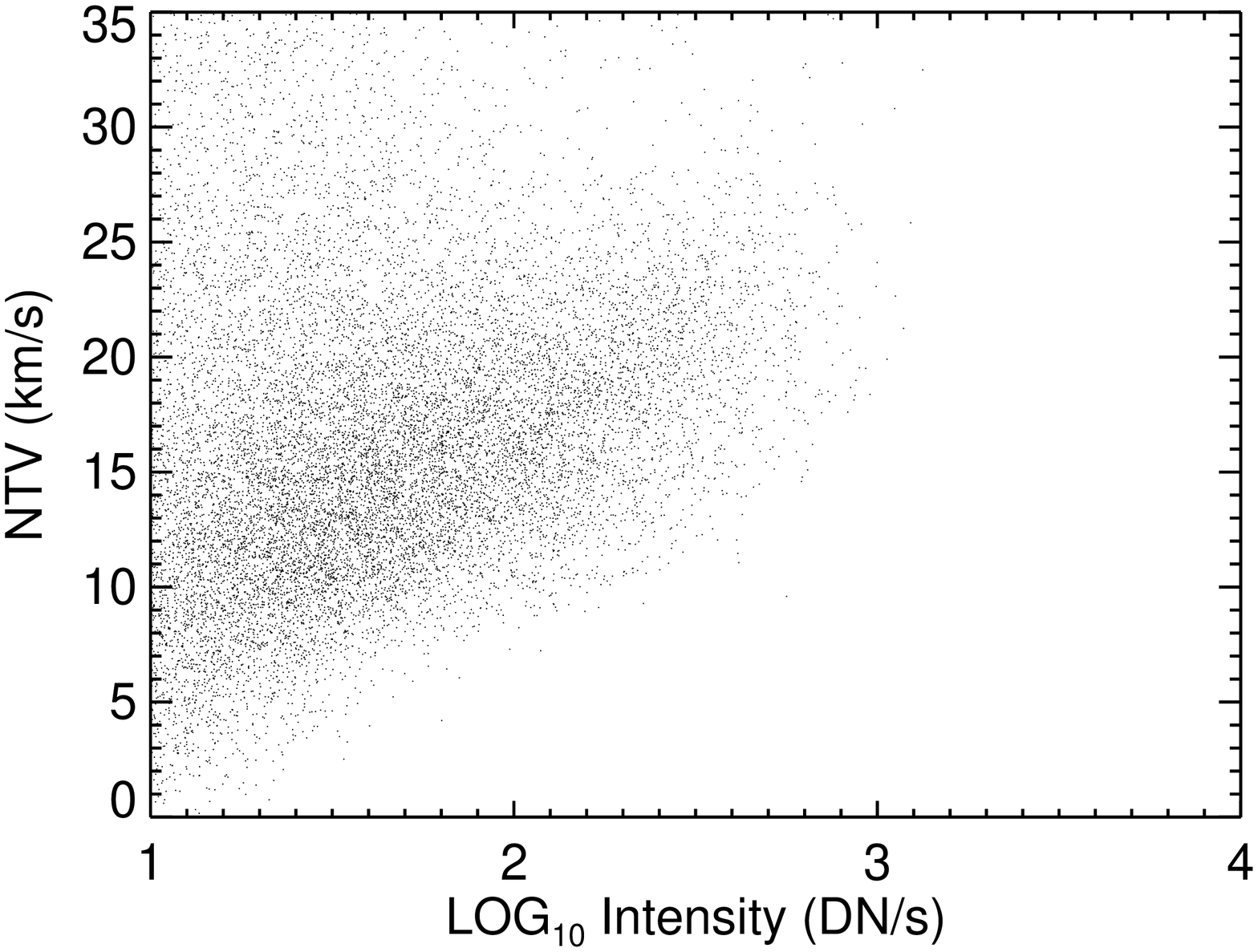}
}
\caption{Left panel: The distribution of non-thermal velocity for the for the QS observed on the 26th September, 2013 (Raster 1) for which the parametric plot is shown in Fig.~\ref{1.5}. Right panel: Scatter plot of the NTVs as a function of the intensity of the \ion{Si}{iv} 1393.75 \AA\ spectral line.}
\label{1.6}
\end{figure*}
\end{center}
\begin{center}
\begin{figure*}
\includegraphics[scale=0.4,angle=90,width=15cm,height=12cm,keepaspectratio]{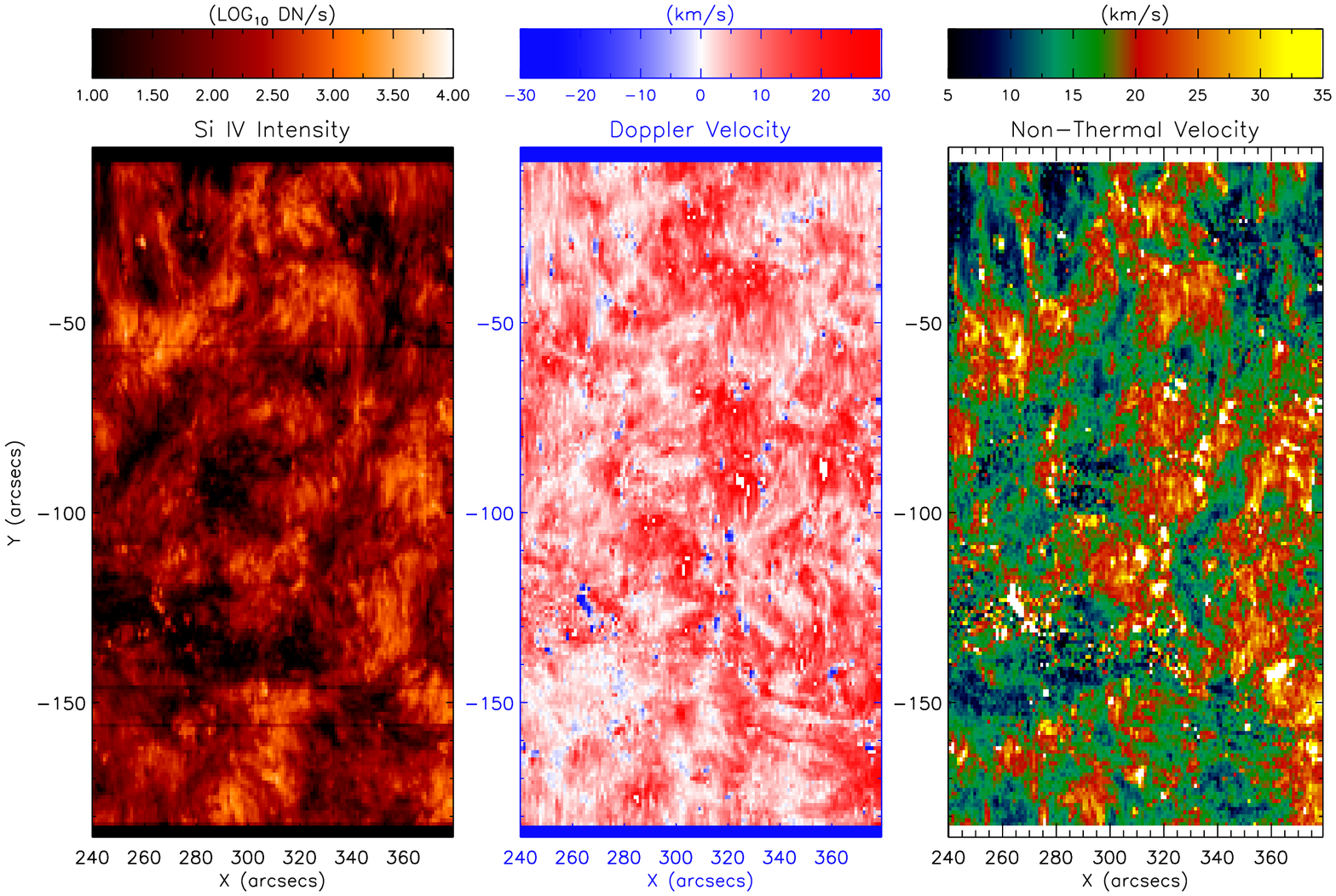}
\caption{Parametric plots of the QS observed on the 22nd October, 2013 near the disc center having an exposure time of 30 s. The left panel shows the intensity of the IRIS \ion{Si}{iv} 1393.75 \AA\ line. The middle and right panel indicates corresponding Doppler velocities and non-thermal velocities. The colour bars are shown above the plots.}
\label{1.7}
\end{figure*}
\end{center}

\begin{center}
\begin{figure*}
\mbox{
\includegraphics[scale=0.4,angle=90,width=8cm,height=10cm,keepaspectratio]{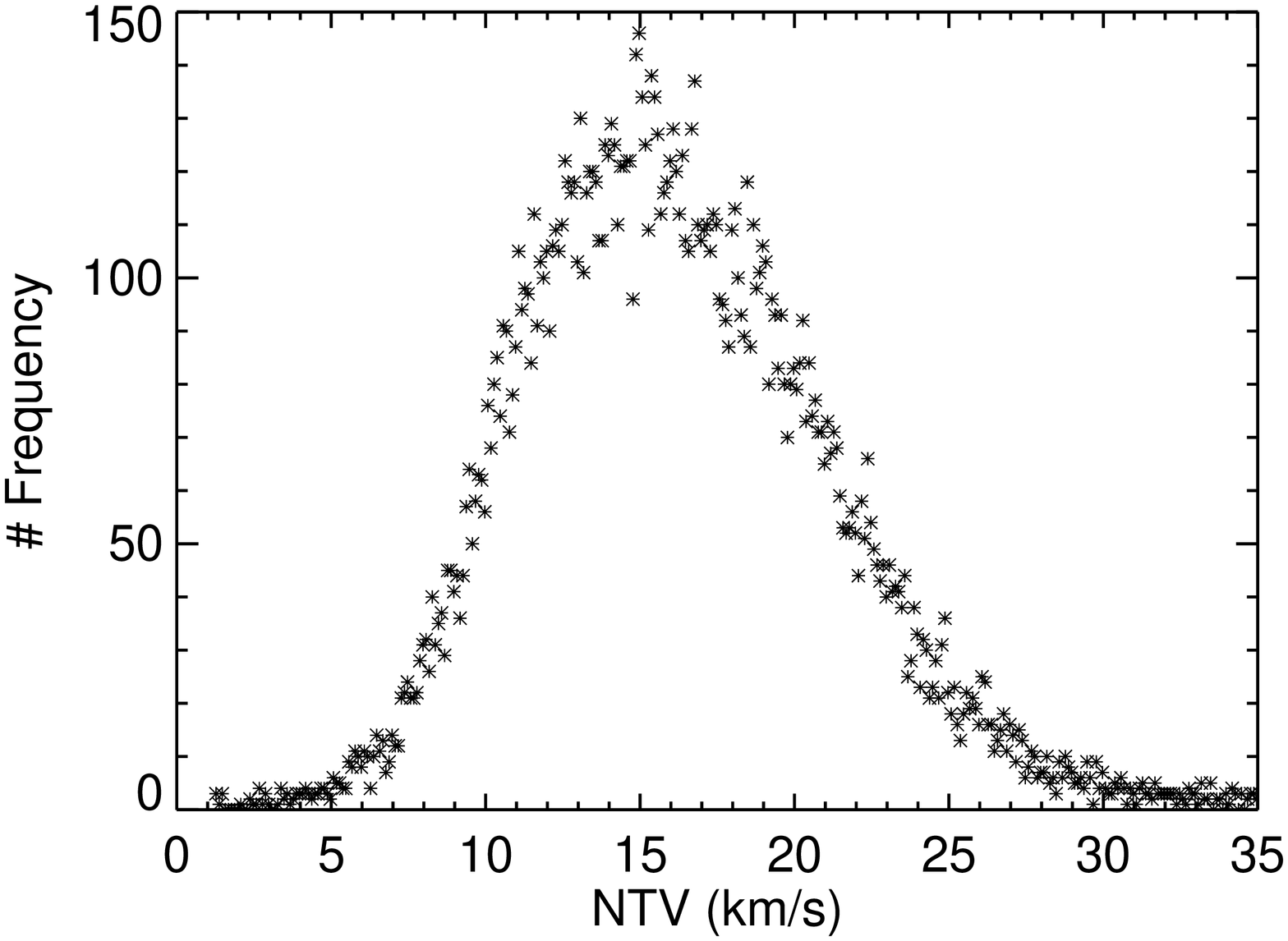}

\includegraphics[scale=0.4,angle=90,width=8cm,height=10cm,keepaspectratio]{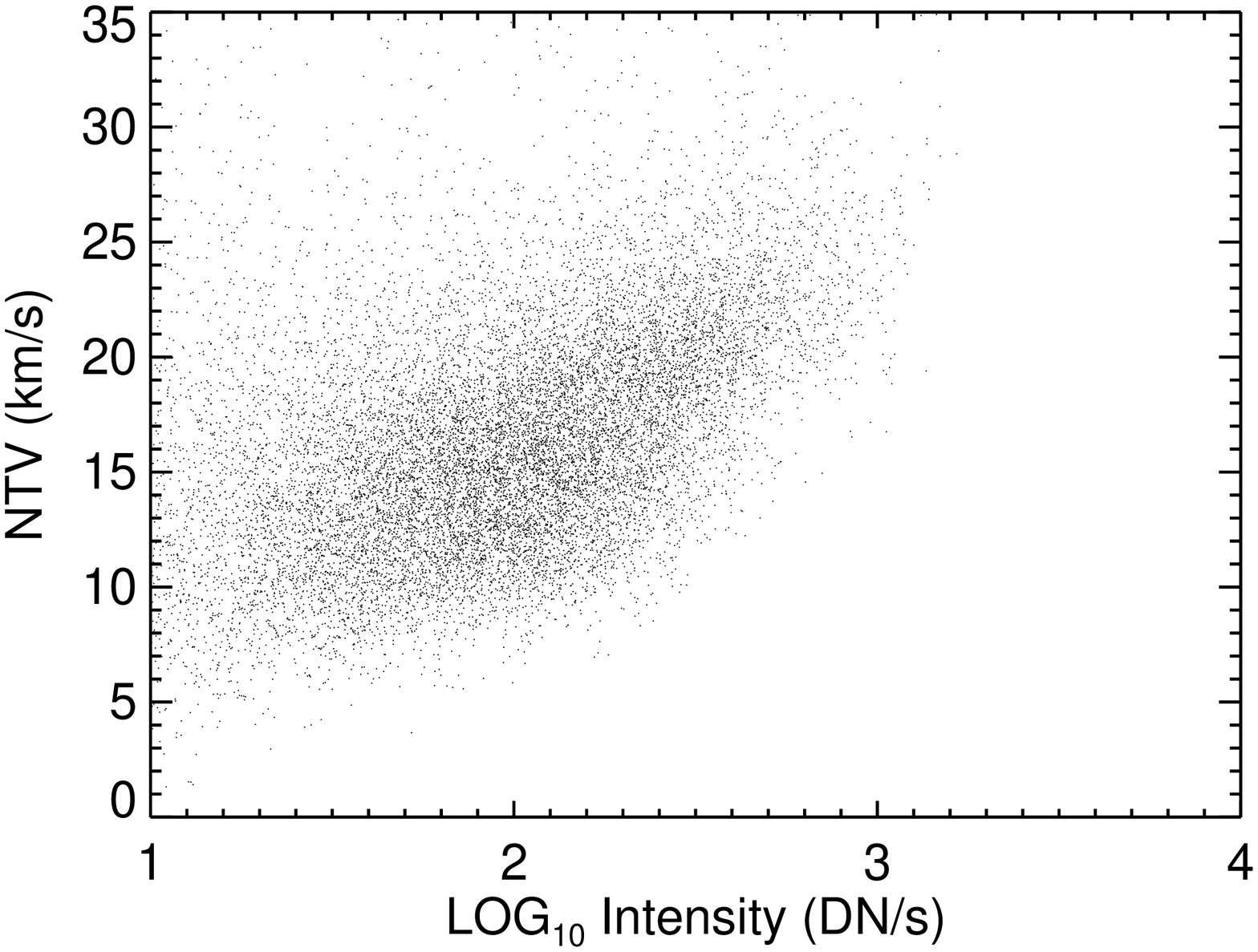}
}
\caption{Left panel: The distribution of non-thermal velocity for the for the QS observed on the 22nd October, 2013 for which the parametric plot is shown in Fig.~\ref{1.7}. Right panel: Scatter plot of NTVs as a function of the intensity of \ion{Si}{iv} 1393.75 \AA\ spectral line.}
\label{1.8}
\end{figure*}
\end{center}

\begin{center}
\begin{figure*}
\includegraphics[scale=0.4,angle=90,width=15cm,height=12cm,keepaspectratio]{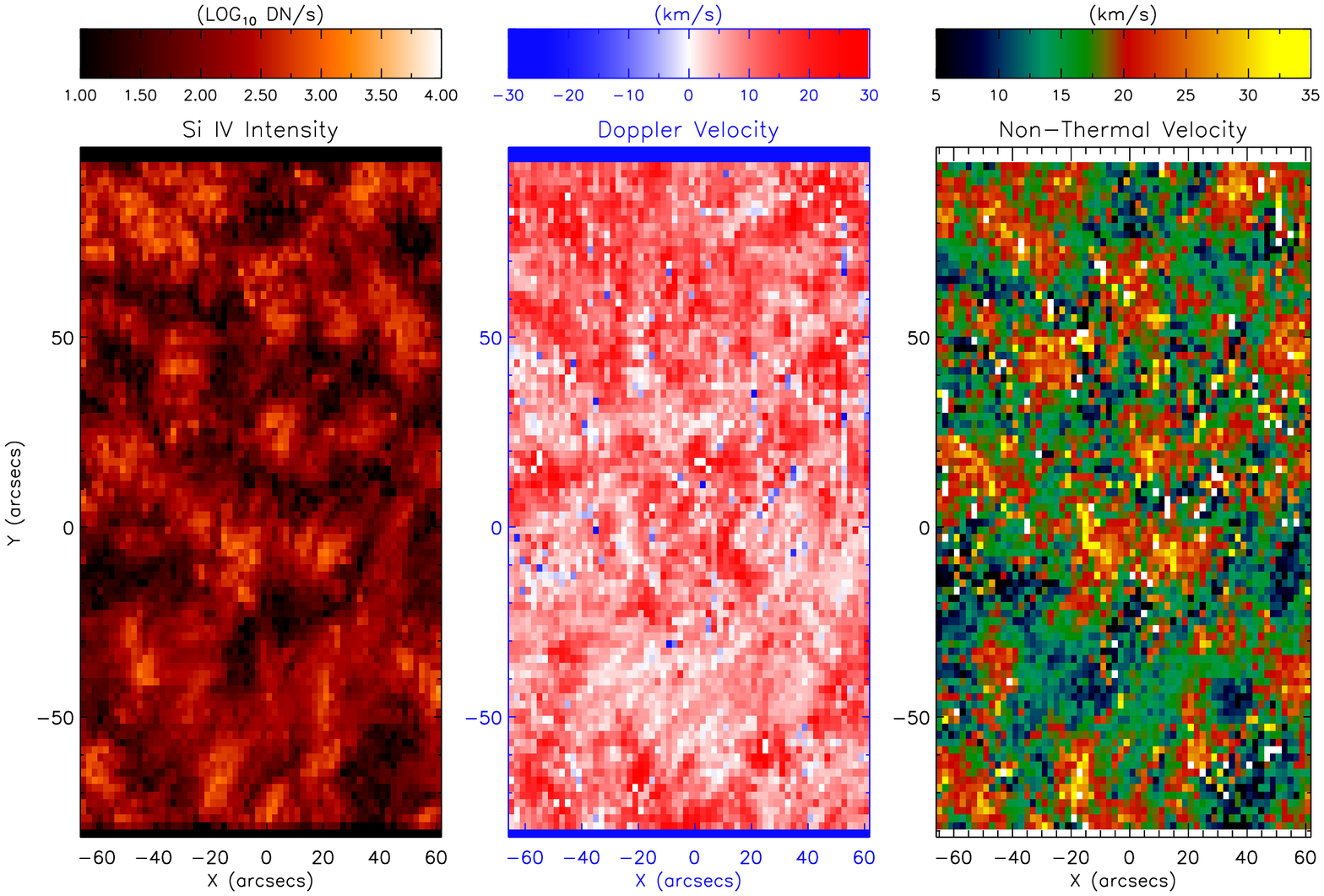}
\caption{Parametric plots of the QS observed on the 27th October, 2013 near the disc center having an exposure time of 4 s. The left panel shows the intensity using the \ion{Si}{iv} 1393.75 \AA\ line observed by IRIS from TR emission. The middle and right panel indicates corresponding Doppler velocities and non-thermal velocities. The colourbars are shown above the plots.}
\label{1.9}
\end{figure*}
\end{center}

\begin{center}
\begin{figure*}
\mbox{
\includegraphics[scale=0.4,angle=90,width=8cm,height=10cm,keepaspectratio]{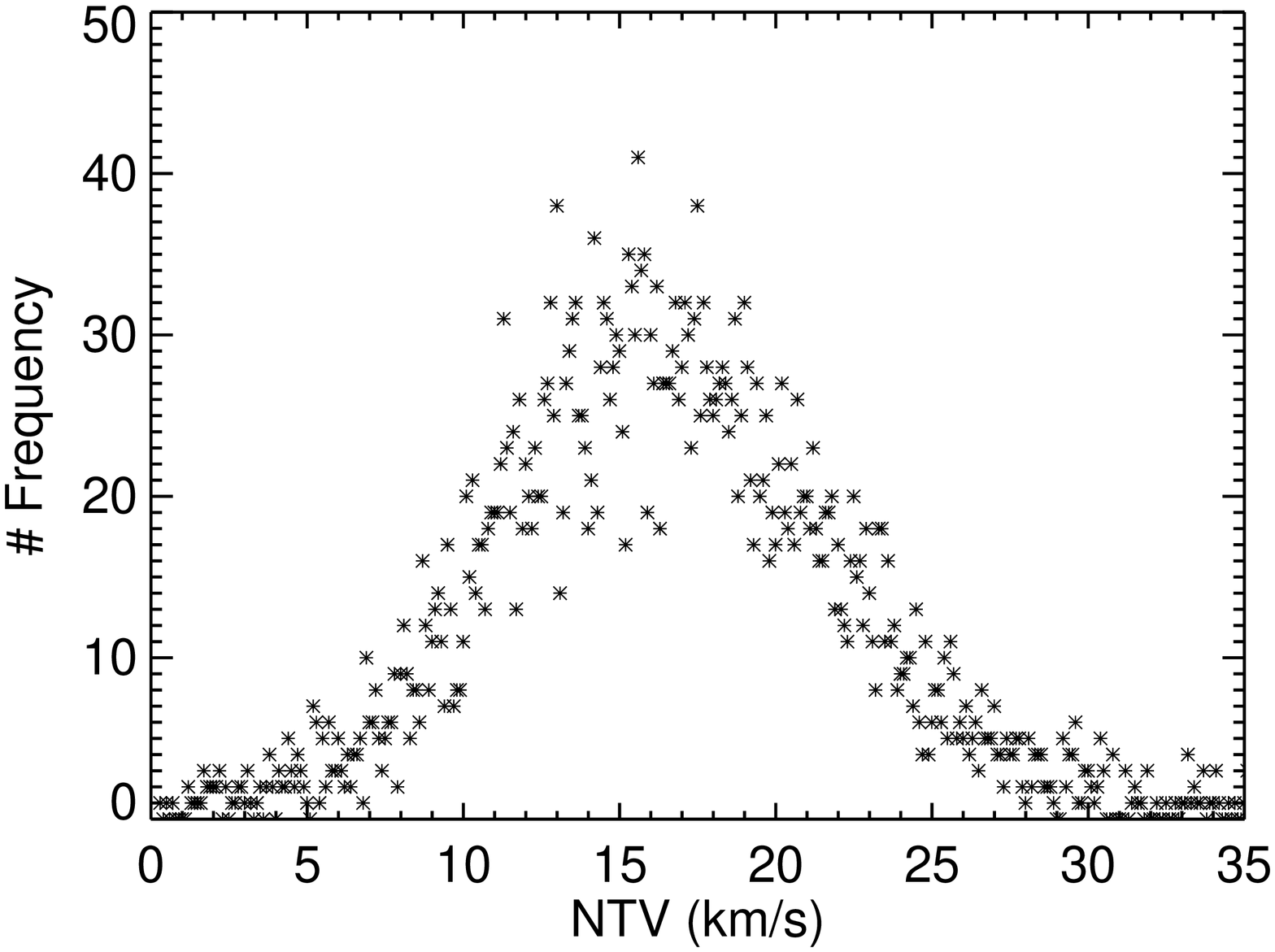}

\includegraphics[scale=0.4,angle=90,width=8cm,height=10cm,keepaspectratio]{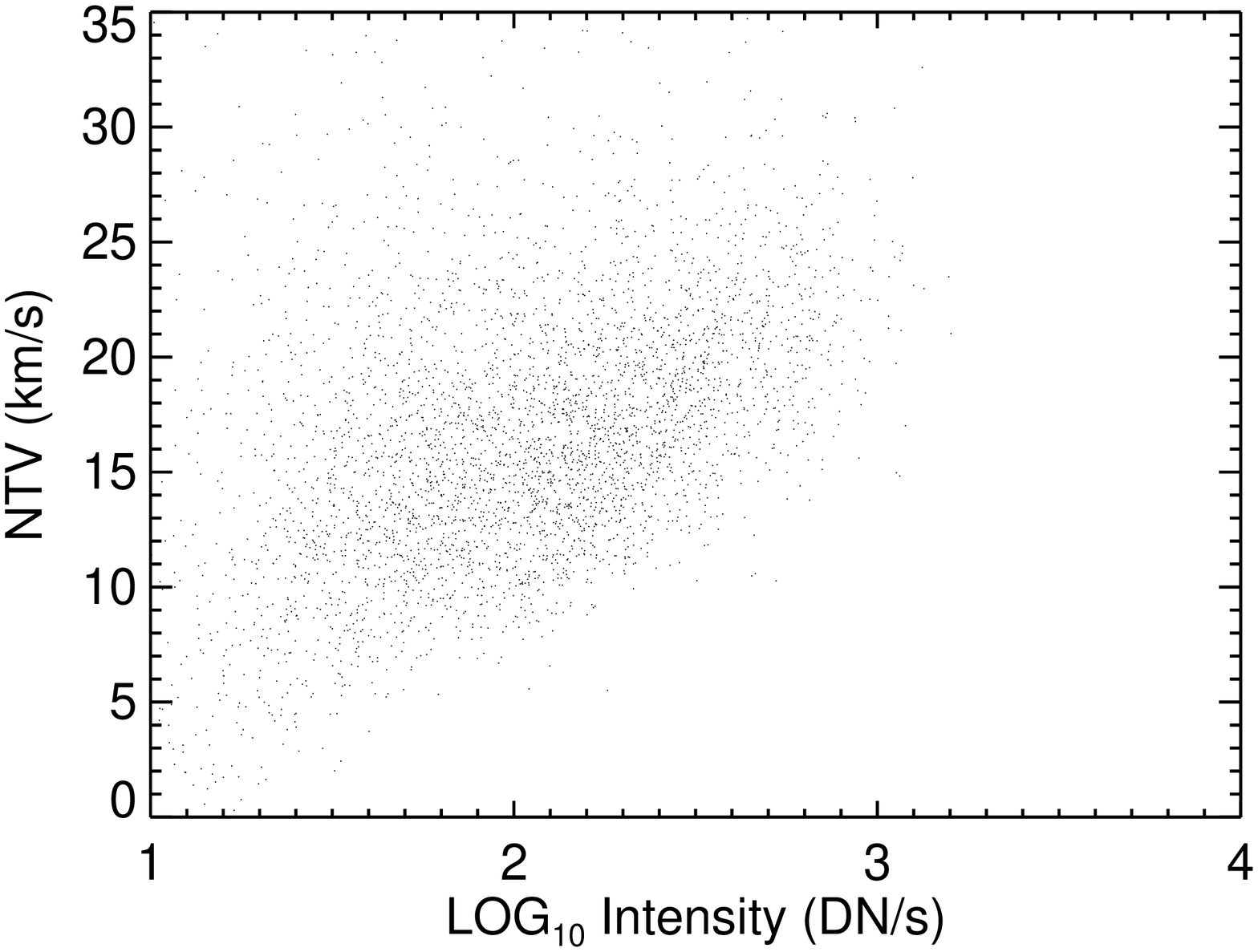}
}
\caption{Left panel: The distribution of non-thermal velocity for the for the QS observed on the 27th October, 2013 for which the parametric plot is shown in Fig. \ref{1.9}. Right panel: Scatter plot of NTVs as a function of the intensity of \ion{Si}{iv} 1393.75 \AA\ spectral line.}
\label{1.10}
\end{figure*}
\end{center}




\begin{center}
\begin{figure*}
\includegraphics[scale=0.4,angle=90,width=15cm,height=12cm,keepaspectratio]{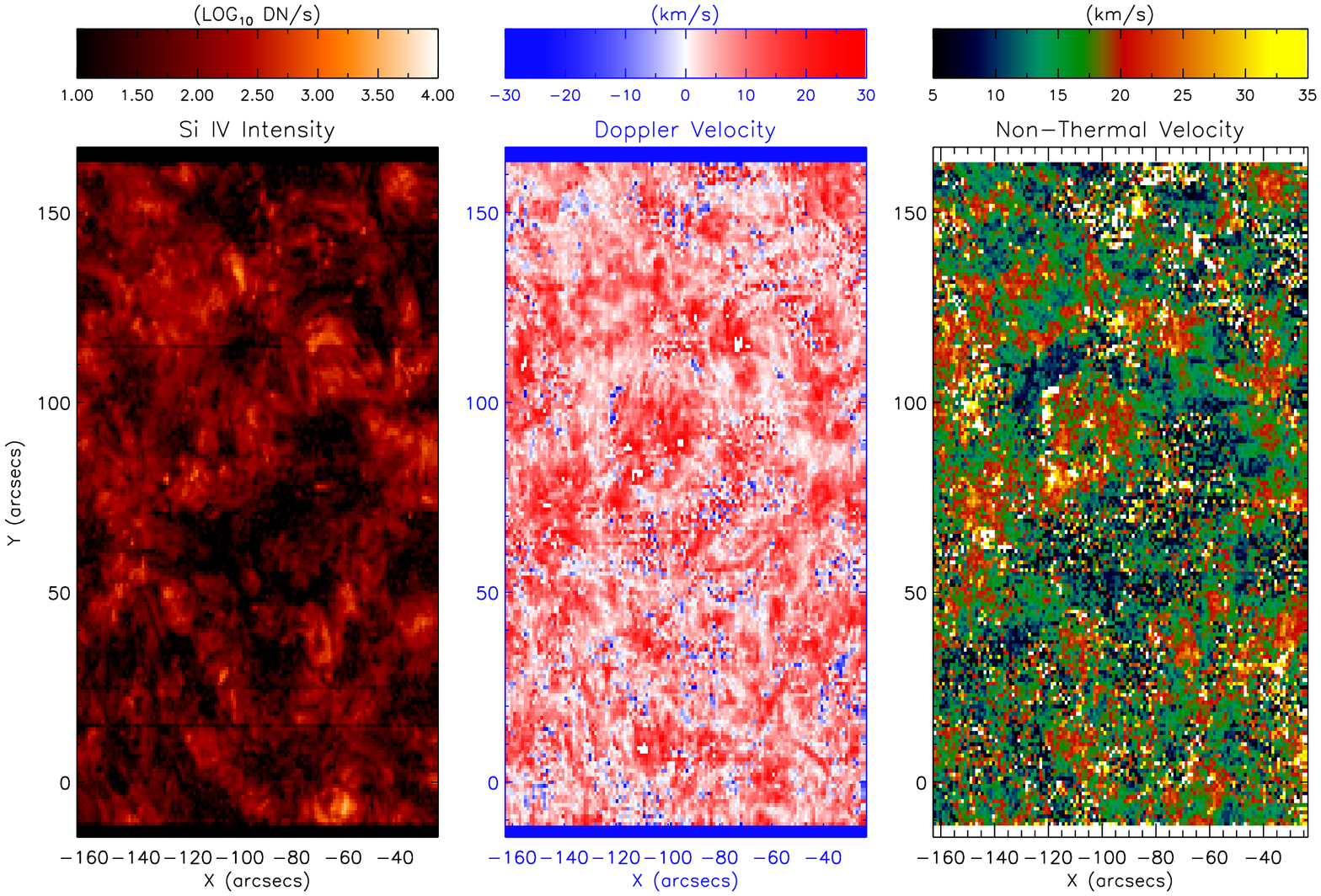}
\caption{Parametric plots of the QS observed on the 25th Febraury, 2014 near the disc center having an exposure time of 8 s. The left panel shows the intensity using \ion{Si}{iv} 1393.75 \AA\ line observed by IRIS from TR emission. The middle and right panel indicates corresponding Doppler velocities and Non-thermal velocities. The colourbars are shown above the plots.}
\label{1.13}
\end{figure*}
\end{center}

\begin{center}
\begin{figure*}
\mbox{
\includegraphics[scale=0.4,angle=90,width=8cm,height=10cm,keepaspectratio]{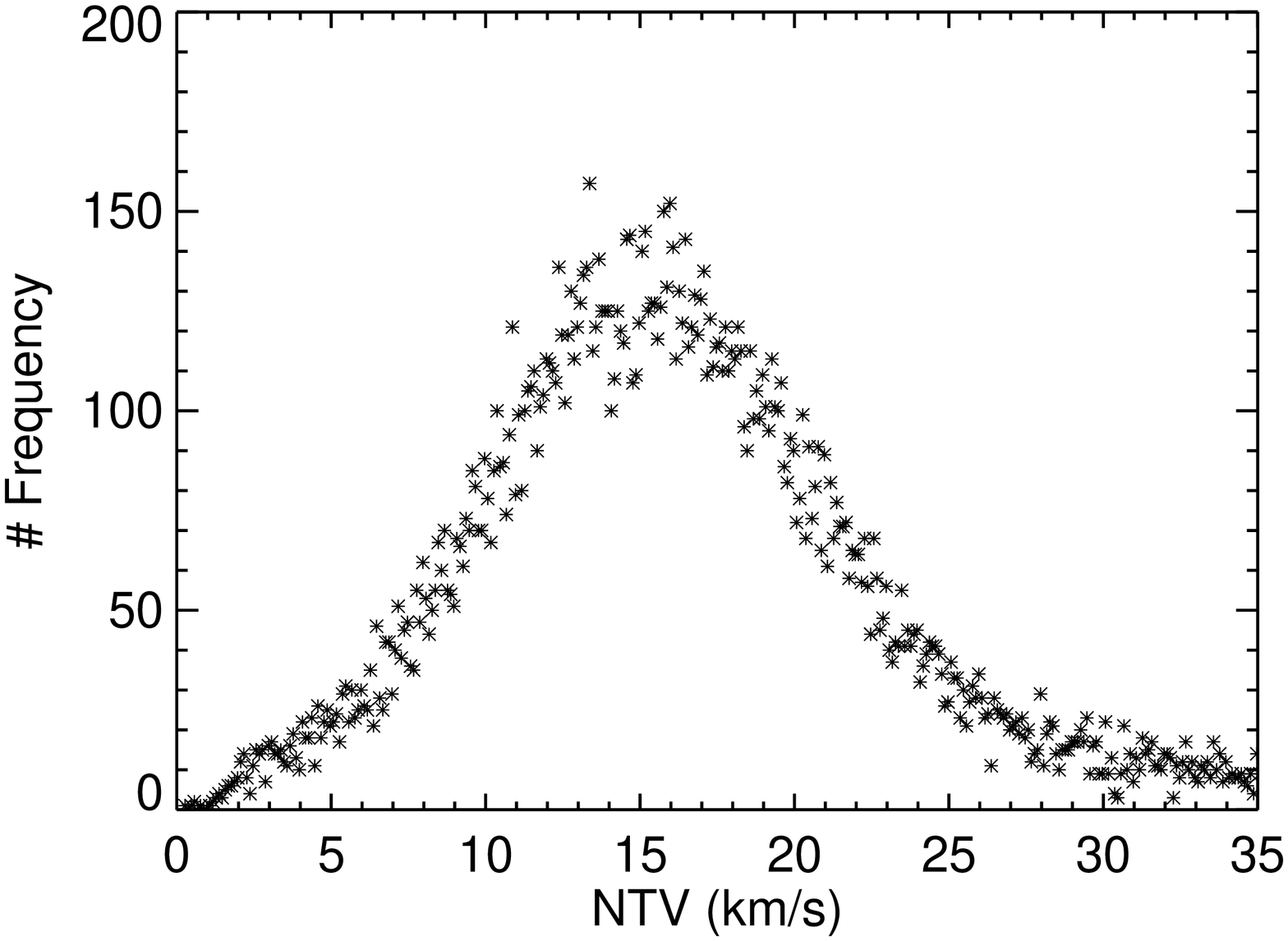}

\includegraphics[scale=0.4,angle=90,width=8cm,height=10cm,keepaspectratio]{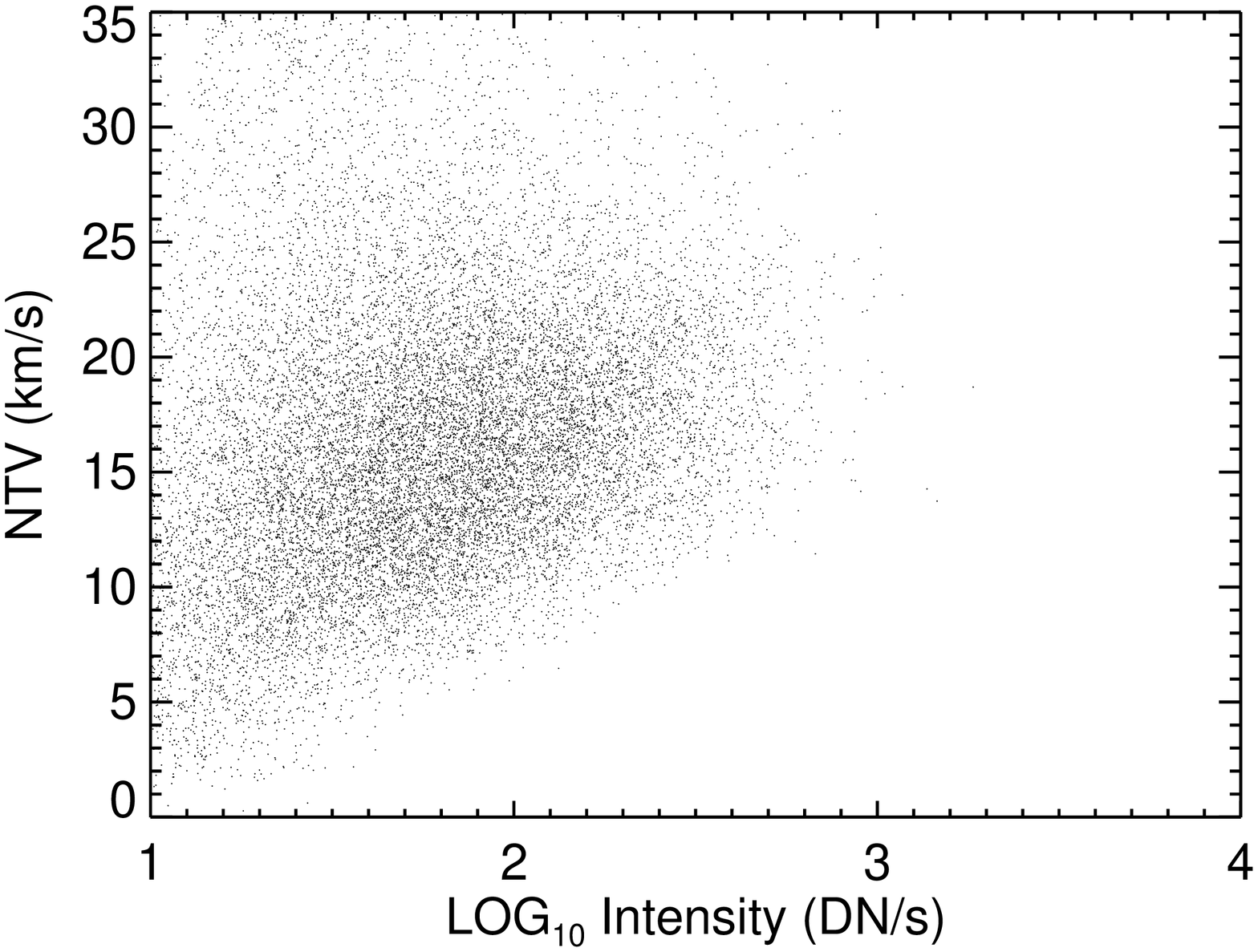}
}
\caption{Left panel: The distribution of non-thermal velocity for the for the QS observed on the 25th February, 2014 for which the parametric plot is shown in Fig. \ref{1.13}. Right panel: Scatter plot of NTVs as a function of the intensity of \ion{Si}{iv} 1393.75 \AA\ spectral line.}
\label{1.14}
\end{figure*}
\end{center}

\begin{center}
\begin{figure*}
\includegraphics[scale=0.4,angle=90,width=15cm,height=12cm,keepaspectratio]{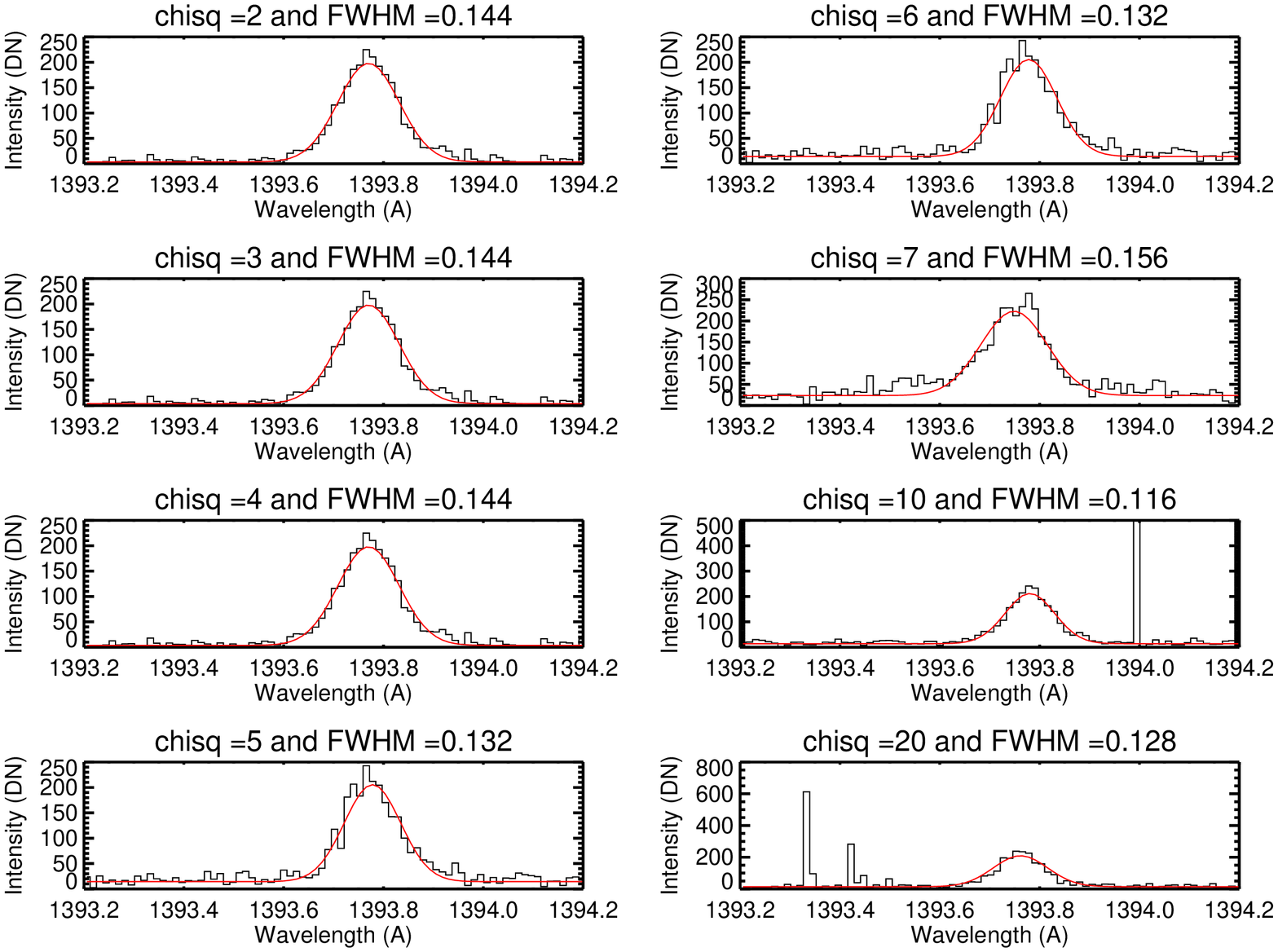}
\caption{Spectral fitting for intensity values of 200 DN and different chi-squared values for the QS observation at the disc center mentioned in the paper.}
\label{spec_fit1.2}
\end{figure*}
\end{center}

\begin{center}
\begin{figure*}
\includegraphics[scale=0.4,angle=90,width=15cm,height=12cm,keepaspectratio]{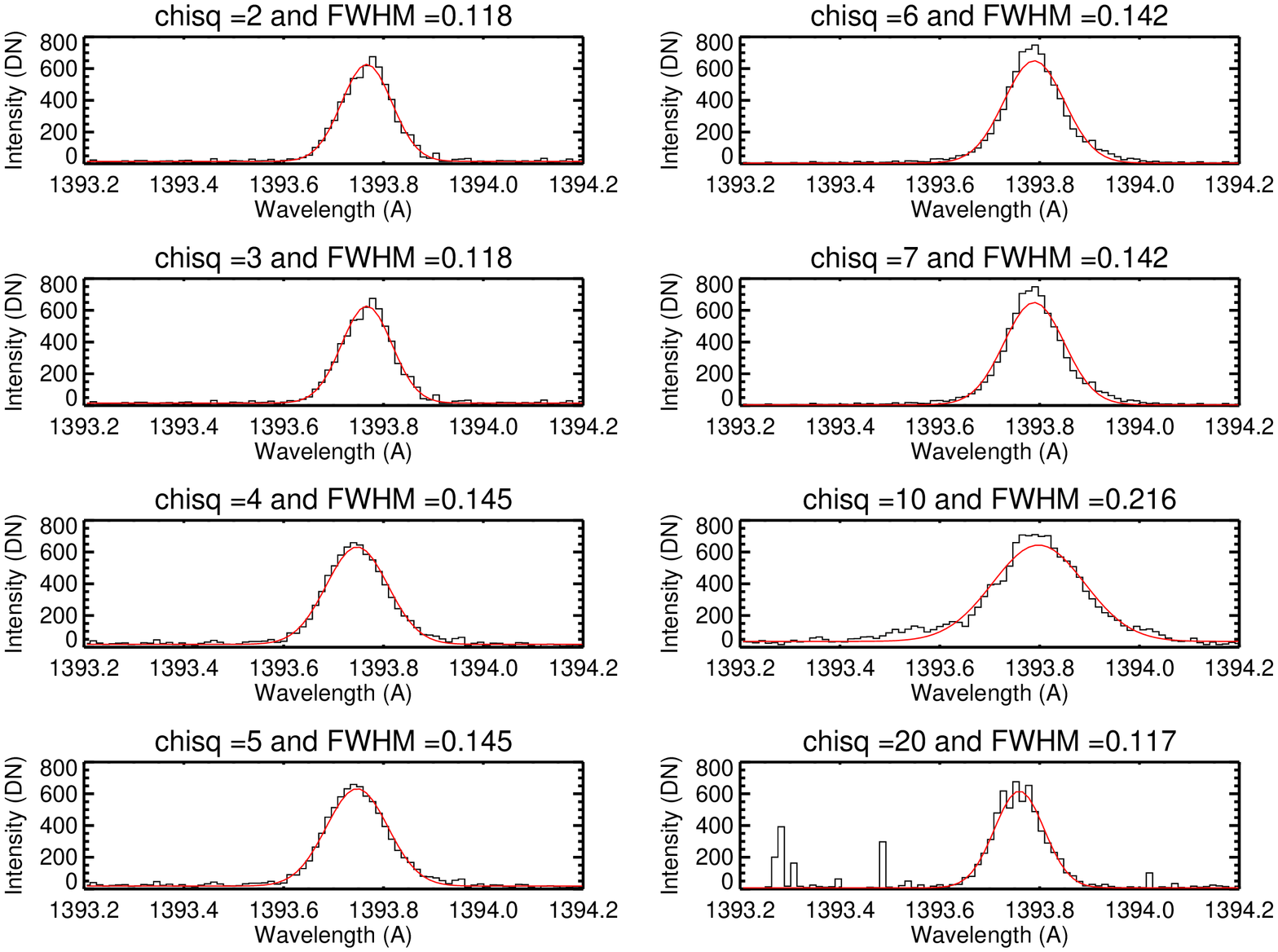}
\caption{Spectral fitting for intensity values of 600 DN and different chi-squared values for the QS observation at the disc center mentioned in the paper.}
\label{spec_fit1.3}
\end{figure*}
\end{center}

\begin{center}
\begin{figure*}
\includegraphics[scale=0.4,angle=90,width=15cm,height=12cm,keepaspectratio]{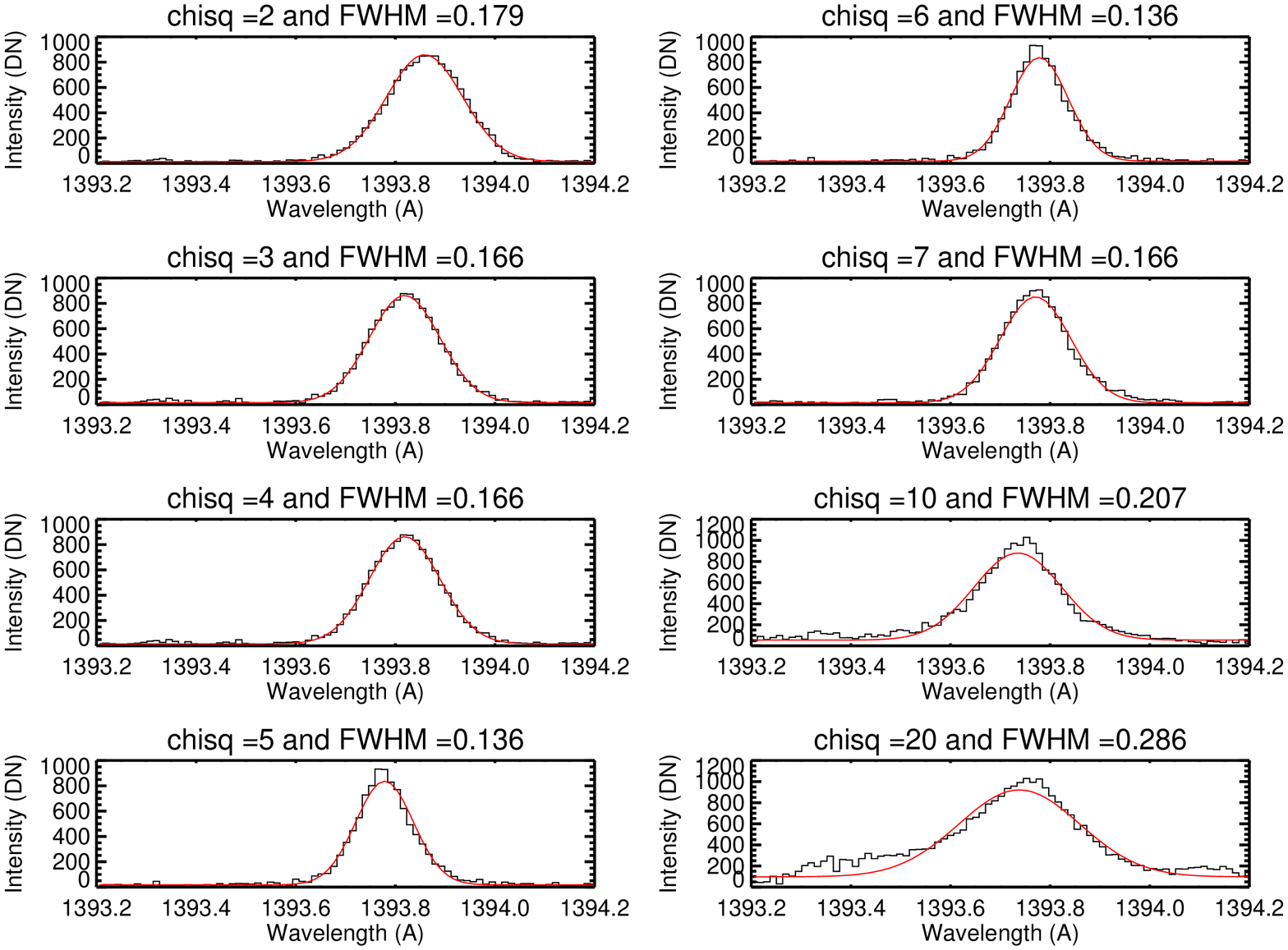}
\caption{Spectral fitting for intensity values of 800 DN and different chi-squared values for the QS observation at the disc center mentioned in the paper.}
\label{spec_fit1.4}
\end{figure*}
\end{center}

\begin{center}
\begin{figure*}
\includegraphics[scale=0.4,angle=90,width=15cm,height=12cm,keepaspectratio]{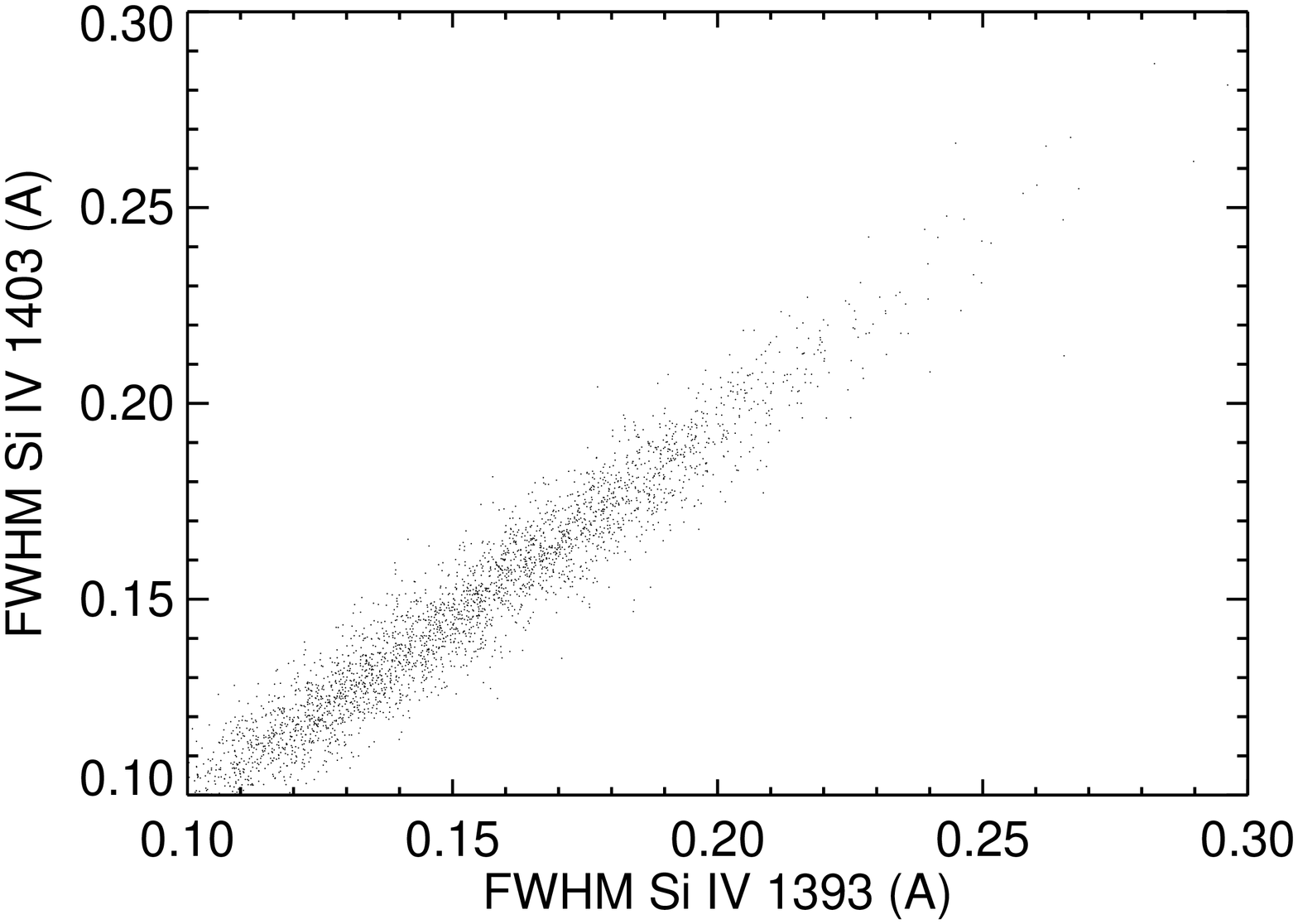}
\caption{Observed FWHM of two different Si IV lines showing strong correlation between two lines}
\label{fwhm}
\end{figure*}
\end{center}

\end{document}